\documentclass[11pt,twoside]{article}

\usepackage{amsmath,amssymb,mathrsfs,tabularx,ragged2e,booktabs,multirow,dcolumn,bigdelim,filecontents,bm,bold-extra}
\usepackage{here,fancybox,arydshln,enumitem,sectsty,stackengine,empheq,cases,braket,caption,subcaption}
\usepackage[hang,flushmargin]{footmisc}
\usepackage[dvipdfmx]{graphicx,color}
\usepackage[dvipsnames]{xcolor}
\usepackage[T1]{fontenc}
\usepackage{fancyhdr,pdflscape,afterpage,capt-of,graphbox,cancel,mwe,ulem,mdframed}
\usepackage{array,dcolumn,threeparttable,makecell,endnotes,xurl,setspace,newtxtext,newtxmath,titlesec,tocloft}

\usepackage[pdfencoding=auto,linkbordercolor=CornflowerBlue,citecolor=tb,linkcolor=tb,pdfpagelabels=false]{hyperref}
\hypersetup{colorlinks=true,allbordercolors=CornflowerBlue}

\usepackage[font={small},labelsep=space]{caption}
\usepackage{subcaption}

\DeclareCaptionLabelFormat{vert}{\textbf{{#1} {#2}~|}}
\usepackage[modulo]{lineno}

\newcommand{\f}[2]{\frac{#1}{#2}}
\newcommand{\ko}[1]{\left( #1 \right)}
\newcommand{\bmt}[1]{{{\mbox{\boldmath$ #1 $}}}}
\newcommand{\q}[1]{`#1'}

\renewcommand*{\thepage}{\footnotesize\arabic{page}}

\def\mtitle{Dynamic development of public attitudes towards science policymaking}
\renewcommand{\baselinestretch}{1.31}

\usepackage[compress,authoryear,round]{natbib}
\setcitestyle{notesep={;},aysep={,},yysep={;}}

\usepackage{etoolbox}
    \newif\ifabbreviation
    \pretocmd{\thebibliography}{\global\abbreviationfalse}{}{}
    \AtBeginDocument{\global\abbreviationfalse}
    \DeclareRobustCommand\acroauthor[2]{%
      \ifabbreviation #2\else #1\global\abbreviationtrue\fi}

\makeatletter
\newcommand*{\myfnsymbol}[1]{\ensuremath{%
\ifcase#1 \or \ast \or \dagger \or \clubsuit \or \heartsuit \else \@ctrerr \fi}}
\makeatother

\definecolor{tb}{rgb}{0.24, 0.43, 0.91}
\renewcommand*{\cite}[1]{\textcolor{tb}{\citep{#1}}}
\newcommand*{\citex}[2]{\textcolor{tb}{\citep[#1]{#2}}}
\newcommand*{\citek}[1]{\textcolor{tb}{\citet{#1}}}
\newcommand*{\citeq}[2]{\textcolor{tb}{\citet[#1]{#2}}}
\newcommand*{\citec}[3]{\textcolor{tb}{\citep[#1][#2]{#3}}}

\renewcommand{\enotesize}{\fontsize{9pt}{10pt}\selectfont}

\begin{document}

\quad
\vspace{-1.4cm}
%%%%%%%%%%%%%
\begin{flushright}
September 2015
\end{flushright}
%%%%%%%%%%%%%

\vspace{0.5cm}

\begin{center}
\fontsize{15pt}{16pt}\selectfont\bfseries
\mtitle
\end{center}

\renewcommand*{\thefootnote}{\textcolor{black}{\myfnsymbol{\value{footnote}}}}

%%%%%%%%%%%%%%%%%%%%%%%%%%
\vspace*{0.8cm}
\centerline{%
{Keisuke Okamura}\,\footnote{\,{\tt okamura@ifi.u-tokyo.ac.jp}}${}^{,}$%
\footnote{\,\href{https://orcid.org/0000-0002-0988-6392}{\tt \textcolor{black}{orcid.org/0000-0002-0988-6392}}}${}^{;\,1}$}

%%%%%%%%%%%%%%%%%%%%%%%%%%
\vspace*{0.6cm}
{\small
\centerline{\textit{
${}^{1}$Ministry of Education, Culture, Sports, Science and Technology (MEXT),}}
\centerline{\textit{
3-2-2 Kasumigaseki, Chiyoda-ku, Tokyo 100-8959, Japan.}}
}
\vspace*{0.5cm}
%%%%%%%%%%%%%%%%%%%%%%%%%%

\vspace{1.5cm}
\noindent\textbf{Abstract.}
\quad
%%%%%%%%%%%%%%%%%%%%%%%%%
Understanding the heterogeneity of mechanisms that form public attitudes towards science and technology policymaking is essential to the establishment of an effective public engagement platform.
Using the 2011 public opinion survey data from Japan ($n=6,136$), I divided the general public into three categories: the Attentive public, who are willing to actively engage with science and technology policymaking dialogue; the Interested public, who have moderate interest in science and technology but rely on experts for policy decisions; and the Residual public, who have minimal interest in science and technology.
On the basis of the results of multivariate regression analysis, I have identified several key predispositions towards science and technology and other socio-demographic characteristics that influence the shift of individuals from one category of the general public to another.
The findings provide a foundation for understanding how to induce more accountable, evidence-based science and technology policymaking.

\vspace{0.8cm}


\vfill
%%%%%%%%%%%%%%%%%%%%%%%%%

\thispagestyle{empty}
\setcounter{page}{1}
\setcounter{footnote}{0}
\setcounter{figure}{0}
\setcounter{table}{0}
\setcounter{equation}{0}

\setlength{\skip\footins}{10mm}
\setlength{\footnotesep}{4mm}

\vspace{-1.6cm}

\newpage
%\linenumbers
\renewcommand{\thefootnote}{\arabic{footnote}}

\setlength{\skip\footins}{10mm}
\setlength{\footnotesep}{4mm}

\let\oldheadrule\headrule
\renewcommand{\headrule}{\color{VioletRed}\oldheadrule}

\pagestyle{fancy}
\fancyhead[LE,RO]{\textcolor{VioletRed}{\footnotesize{\textsf{\leftmark}}}}
\fancyhead[RE,LO]{}
\fancyfoot[RE,LO]{\color[rgb]{0.04, 0.73, 0.71}{}}
\fancyfoot[LE,RO]{\scriptsize{\textbf{\textsf{\thepage}}}}
\fancyfoot[C]{}

\newpage
\tableofcontents
\thispagestyle{empty}

%\linenumbers

\vfill

%%%%%%%%%%%%%%%%%%%%%%%%%%
\begin{mdframed}[linecolor=gray]
\begin{tabular}{l}
\begin{minipage}{0.11\hsize}
\hspace{-2.5mm}\includegraphics[width=1.8cm,clip]{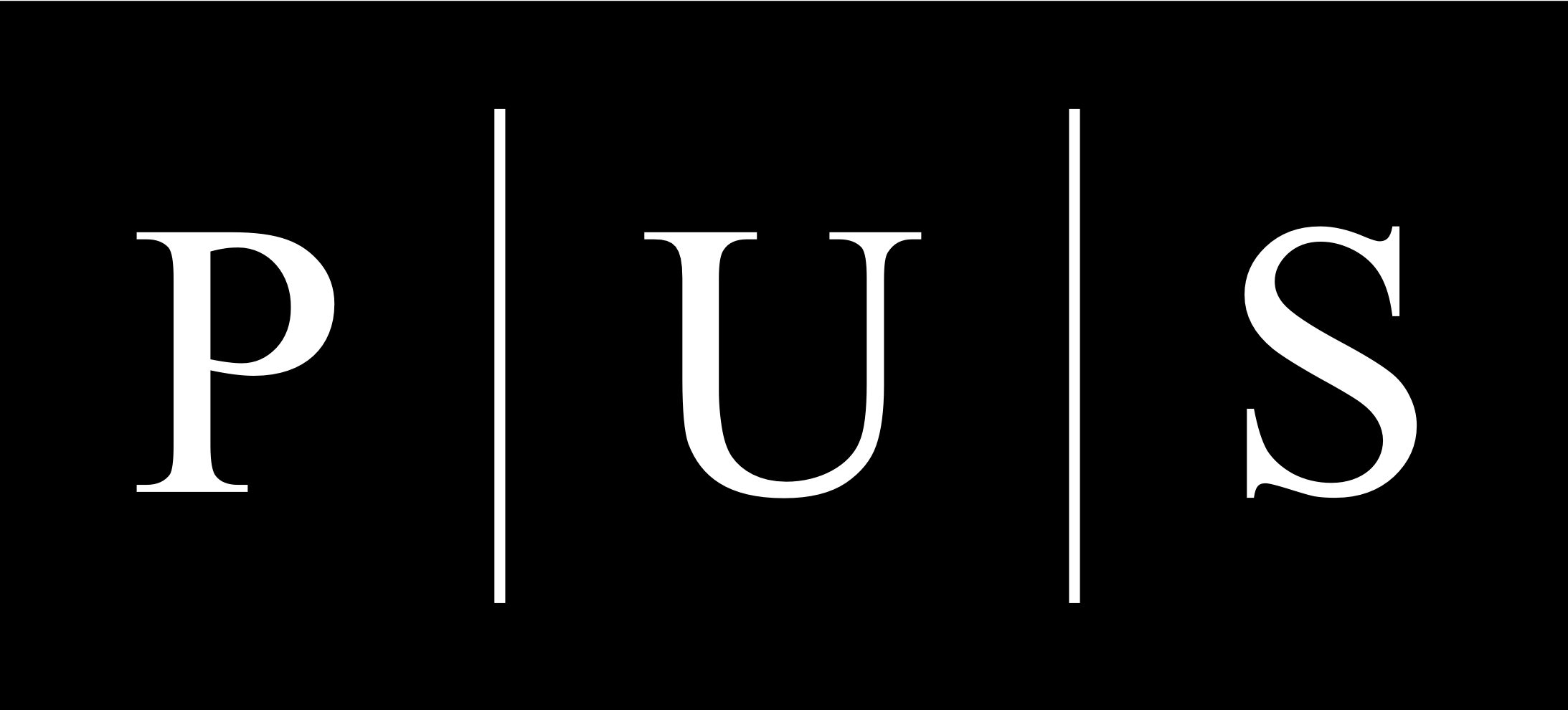}
\end{minipage}
\begin{minipage}{0.86\hsize}
\setstretch{0.9}
\textcolor{gray}{\footnotesize\textsf{%
This is the accepted version of a paper published in \textit{Public Understanding of Science}, {25}(4):465--479, 2016; \href{https://doi.org/10.1177/0963662515605420}{DOI:10.1177/0963662515605420}. 
The Online Appendix accompanying the published version of the paper has been seamlessly integrated into this single file as \q{Supplementary Materials}.}}
\end{minipage}
\end{tabular}
\end{mdframed}
%%%%%%%%%%%%%%%%%%%%%%%%%%

\newpage
%%%%%%%%%%%%%%%%%%%%%%%%%%%%%%%%%
%%%%%%%%%%%%%%%%%%%%%%%%%%%%%%%%%
\section{Introduction}
%%%%%%%%%%%%%%%%%%%%%%%%%%%%%%%%%

In contemporary societies, a significant emphasis is placed on promoting strong science communications, that is, communication links between the mediators of science and technology (S\&T) and society in general \cite{Bauer:2007,Bauer:2011,Gaskell:2005,Jacobson:2004,Jensen:2008,Nature:2004,Nisbet:2009,Scheufele:2014}.
Japan is no exception to this trend.
Reflecting on numerous lessons learned from the Great East Japan Earthquake and the subsequent Fukushima nuclear power plant accident in March 2011,\endnote{%
For detailed information about the facts and circumstances of the Great East Japan Earthquake and the nuclear disaster, see the website of the \citeq{~since 11 March 2011}{GEJE}.}
the country's latest S\&T Basic Plan states,

%%%
\begin{quote}
It is necessary for the government to obtain the public understanding, trust and support in formulating and promoting S\&T and innovation policies by deeply and accurately grasping the expectations and concerns of the general public and society [with regard to the policymaking process]. \cite{4thS&T}
\end{quote}
%%%

Indeed, common recommendations on how to deal with the challenge of post-disaster science communication in Japan have strongly suggested that the link between S\&T and civil society should be strengthened by narrowing the gap between S\&T experts and the general public \citec{\textcolor{black}{e.g.}\/}{}{MEXT:2012}.

However, how much do we actually know about those to whom we refer as the \q{public}?
Who is \q{the general public}?
Most discussions of how science communications can be effectively implemented appear to be based on a presupposition that the general public is more or less homogeneous in how it responds to changes in circumstances related to S\&T and the policymaking process.
However, reality appears more complex and quite often ambiguous.
For instance, some people believe that increased public distrust in scientists, which has actually been observed in Japan since the disaster \citex{\textcolor{black}{~see also Figure \ref{figS1}}}{MEXT:2012}, should increase public willingness to actively engage in the policymaking process since people would no longer like to delegate policymaking to the experts, whereas others see the opposite effect, with an anticipation that people would simply lose their interest in S\&T in general.
Which is true?

Probably both claims are partly true, depending on which part of the \q{public} one is referring to, as one can identify different subgroups of society, each of which shares common predispositions about S\&T or socio-demographic characteristics.
As S\&T today is more closely connected to society than ever before, the style used to communicate with different members of society should be correlated with their particular needs and attitudes towards S\&T \cite{Dijkstra:2012b,Dijkstra:2012a}.
Thus, obtaining a clearer and more refined understanding of the potentially heterogeneous views and behavioural responses of the public is crucial for deriving useful policy implications aimed at improving science communications and the governance structure.

To explore these issues, in this study, I investigate the case of Japan in 2011, the year in which it suffered the major earthquake, tsunami and nuclear disaster, when the need for more effective and accountable policymaking processes was strongly recognised in society.

%%%%%%%%%%%%%%%%%%%%%%%%%%%%%%%%%
%%%%%%%%%%%%%%%%%%%%%%%%%%%%%%%%%
\section{Heterogeneous characteristics of the general public}
%%%%%%%%%%%%%%%%%%%%%%%%%%%%%%%%%

My hypothesis regarding the heterogeneity of attitudes towards S\&T policymaking within the general public relies on a variation of the traditional \q{stratified pyramid model} first used to illustrate American foreign policy \cite{Almond:1950} and later applied in the context of S\&T policy formation \cite{Miller:1983a,Miller:1983b}.
Here, the definition of the strata is slightly modified to account for Japan's specific social structure.

The first two levels of the pyramidal structure are decision makers, including politicians and high-level government officials, and policy leaders, who provide the decision makers with scientific and technical advice.
The third level is the Attentive public; this level represents those members of the general public who actively follow S\&T issues professionally or personally.
These people are expected to have expertise and/or a strong, consistent interest in S\&T issues, as well as an intention to engage with policymaking dialogue.
The fourth level is the Interested public; this level comprises people who passively follow S\&T issues but accept the decisions made by active players.
They have some interest in S\&T issues, although they tend to rely on experts when it comes to policymaking. 
At the base of the pyramid is the Residual public; these people are relatively inattentive to and uninterested in S\&T issues, let alone policy formation.

Based on this model, the first question to be answered is as follows: what are the key socio-demographic characteristics and predispositions towards S\&T that divide the general public into each of the three categories indicated in the policymaking pyramid (Attentive, Interested and Residual)?

%%%%%%%%%%%%%%%%%%%%%%%%%%%%%%%%%
\subsection{The National Institute of Science and Technology Policy survey data}
%%%%%%%%%%%%%%

To answer this question, I used data derived primarily from the National Institute of Science and Technology Policy \citep{NISTEP-report:2012} of Japan.\endnote{%
Detailed description of the survey design can be found in the NISTEP Research Material \cite{NISTEP-report:2012} and the codebooks attached to the original data \cite{NISTEP-data:2012}.}
As part of its mission to develop data infrastructure, NISTEP conducted an Internet survey on public attitudes towards S\&T every month from November 2009 to March 2012.
To collect the data, NISTEP contacted Internet survey companies that already had a pool of potential participants.
These companies distributed to prospective participants an email describing the survey and inviting their participation.
Individuals who agreed to participate received another email message containing instructions for answering the web-based questionnaire and were asked to visit the survey website.
The website contained an introductory page providing information about the study.
This information included details on the study's purpose, a consent statement and an explanation that the survey was not a test and that the respondent should answer each question in a straightforward and honest manner.

The contents of the questionnaire varied across different months of the survey period, with some modifications being made to the wording of certain questions and other questions being added or removed according to the relevant data requirements at the time.
The biggest change to the survey content was the addition of new questions in April 2011 just after the earthquake and the subsequent Fukushima nuclear power plant accident in March 2011; these questions form the basis for the analysis in this study.
Specifically, the data used throughout this article were collected from April 2011 to November 2011.
The dataset and codebooks were taken from the \citek{NISTEP-data:2012} website and translated from Japanese to English where necessary.
The data in each of the eight data files corresponding to the eight months of the survey were subsequently examined and cleaned by removing irrelevant or nonsensical responses as well as recoding or reclassifying variables where appropriate.
An aggregate dataset was then created by combining the eight data files, which consisted of samples from all areas of Japan, including the affected areas by the earthquake (see Table \ref{tab:month-region}).
The total number of respondents was 6,136.

The socio-demographic profiles of the aggregate sample population are summarised in Table \ref{tab:demog}.
\q{Gender} is a dichotomous variable with males coded as 0 and females coded as 1 (50\% female).
\q{Age} is measured as a continuous variable in years ($M=40.0$ years, $\text{standard deviation (\textit{SD})}=16.0$ years).
\q{Education} is a dichotomous variable representing respondents' educational attainment; this variable was coded as 1 if the respondent had attained higher education (e.g.\ college- or university-level education) and as 0 if the respondent had educational attainment lower than this level (45\% highly educated).
Respondents were also given the chance to select \q{other} in response to this question.
However, these responses were treated as missing values to avoid response bias and to ensure proper statistical analysis.
\q{Marital status} is also a dichotomous variable, coded as 1 if a respondent was married and 0 otherwise (57\% married).
\q{Income} represents the total annual household income (after tax) and was measured using 14 categories; these ranged from 1 (under \textyen{1,000,000}) to 14 (\textyen{20,000,000} or greater).
The median of the valid sample was category 5, which represents the income bracket \q{\textyen{4,000,000}--\textyen{4,999,999}}.

In collecting the data for each month in the survey period identified, NISTEP used a stratified sampling method by gender and age group (15--19, 20--29, 30--39 and so on up to 60--69) in an attempt to recruit approximately the same number of respondents (at least 60 persons) in each stratum \cite{NISTEP-report:2012}.
As such, individuals had an unequal probability of being selected, and the respondents were not the same persons in different survey months.
Therefore, to ensure that survey findings were relevant to the Japanese population, I computed and assigned sampling weights to each $\mathrm{gender}\times\mathrm{age}$ stratum.
The demographic information needed to take the finite population corrections into account was obtained from the population estimates indicated by a national census \cite{SBJ:2012}.
In all statistical analyses conducted below, it was ensured that the sampling weights were properly accounted for, thereby compensating for the sample design that may have over- or under-represented various segments within the population.

%%==================================================
\begin{table}[!tp]
%\vspace*{2em}
\centering
  \begin{threeparttable}
\caption{\textbf{Results from cluster analysis with respect to attitudes towards S\&T policymaking.}}
\label{tab:clusters}
{\small
    \begin{tabular}{l@{\hspace{0.9cm}}r@{\hspace{0.4cm}}r@{\hspace{0.3cm}}r@{\hspace{0.3cm}}r@{\hspace{0.3cm}}r@{\hspace{0.4cm}}r@{\hspace{0.3cm}}r@{\hspace{0.3cm}}r@{\hspace{0.3cm}}r@{\hspace{0.4cm}}r@{\hspace{0.9cm}}r@{\hspace{0.4cm}}r}
    \toprule
          & \multicolumn{2}{c}{{\sl \bfseries Attentive}} &&& \multicolumn{2}{c}{{\sl \bfseries Interested}} &&& \multicolumn{4}{c}{{\sl \bfseries Residual}} \\ \cline{2-3} \cline{6-7} \cline{10-13}\\[-1em]
Dimension  & \multicolumn{2}{c}{Cluster I} &&& \multicolumn{2}{c}{Cluster {II}} &&& \multicolumn{2}{c}{Cluster {III}} & \multicolumn{2}{c}{Other}\\
          & \multicolumn{2}{c}{($n=986$)} &&& \multicolumn{2}{c}{($n=2,465$)} &&& \multicolumn{2}{c}{($n=1,268$)} & \multicolumn{2}{c}{($n=1,417$)} \\
    \midrule
    {Interest in S\&T}	& 3.54  & (0.65) &&& 2.86  & (0.73) &&& 2.75  & (0.62) & 2.57  & (0.81)\\
    {Participation of the general public} 	& 3.76  & (0.43) &&& 2.93  & (0.58) &&& 2.71  & (0.54) & 3.09  & (0.52)\\
    {Delegation to experts} & 1.94  & (0.78) &&& 3.14  & (0.35) &&& 1.86  & (0.35) & 2.83  & (0.65)\\
    \bottomrule
    \end{tabular}
\vspace{0.5em}
\begin{tablenotes}
\linespread{1.1}
{\footnotesize
\item S\&T: science and technology.
\item Cluster I is characterised by high scores on \q{interest in S\&T} and \q{participation of the general public}, whereas Cluster II is characterised by high scores on \q{delegation to experts}; these traits reflect the Attentive and the Interested classifications, respectively.
Based on the policymaking pyramid definition, the remaining respondents, including both Cluster III and those indicated as \q{other}, together form the Residual classification.
Mean scores of each dimension variable are shown with standard deviations in parentheses.
}
\end{tablenotes}
}
  \end{threeparttable}
\vspace*{1em}
\end{table}
%%==================================================

%%%%%%%%%%%%%%%%%%%%%%%%%%%%%%%%%
\subsection{Cluster analysis}
%%%%%%%%%%%%%%

From the dataset, the three observed variables used to infer respondents' attitudes towards S\&T policymaking are as follows.
First, \q{interest in S\&T}, corresponding to dimension 1, was used to measure the respondents' level of interest in S\&T news and issues; responses were recorded on a 4-point Likert scale from 4 (very interested) to 1 (not interested at all).
For dimension 2, which is associated with \q{participation of the general public}, respondents were asked to what extent they agreed with the statement that \q{where research and development (R\&D) has a significant social impact, the general public should participate in some way in making decisions about R\&D}.
For dimension 3, which is associated with \q{delegation to experts}, the statement presented was that \q{decisions about the direction of R\&D in S\&T should be made by trained and experienced experts who are highly knowledgeable}.
Answers to both questions were recorded on a 4-point Likert scale from 4 (strongly agree) to 1 (strongly disagree), measuring the extent to which people agreed with the two distinct attitudes (i.e.\ \q{participation} and \q{delegation}) towards involvement in the S\&T policymaking process.

On the basis of the responses to these three questions (see Table \ref{tab:keyQscicom} for summary statistics), cluster analysis was applied to the dataset\endnote{%
The segmentation of the sample into clusters was performed using non-hierarchical clustering with squared Euclidean distances and the $K$-means method.}
and three clusters were identified that displayed significantly different ($p<.001$ on Pearson's chi-squared test) baseline profiles with respect to all three dimensions (Table \ref{tab:clusters}).
They can be described as follows: Cluster I (16\%; $n=986$) has a high level of interest in S\&T and is also willing to participate in the policymaking process, Cluster II (40\%; $n=2,465$) has moderate interest in S\&T but prefers to rely on experts when it comes to policymaking, and Cluster III (21\%; $n=1,268$) has a relatively low level of interest in S\&T and no strong preference about who engages with S\&T policymaking.
Clusters I and II were straightforwardly identified as the Attentive and Interested categories of the general public, respectively, in accordance with the aforementioned definitions in the pyramid model.
Cluster III was then combined with the outliers, described as \q{other}, to identify the third category, that is, the Residual group.

%%%%%%%%%%
\begin{figure}[!tp]
\centering
\vspace{-1em}
\noindent
\includegraphics[align=c, scale=1.0, vmargin=1mm]{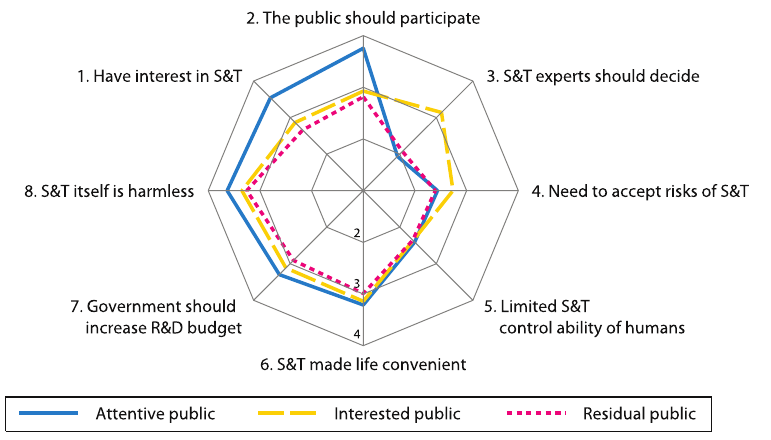}
%\vspace{6mm}
\caption{\textbf{Characteristics and predispositions relating to each cluster of the general public.}
}
\label{fig1}
\vspace*{1em}
\end{figure}
%%%%%%%%%%

Some of the socio-demographic characteristics associated with each of the three general public categories were subsequently examined.
The Attentive group was observed to contain predominantly men (67\%; $p<.001$) and more individuals with higher levels of education (59\%; $p<.001$) than the other two groups.
In contrast, there was no statistically significant difference between the Interested and Residual groups with regard to gender ($p=.307$) or education ($p=.078$), although for education the statistical significance was close to the decision threshold ($p=.05$).
With respect to age, the Interested group contained relatively younger individuals on average ($\text{mean age}=41.9$ years) than the Attentive ($45.2$ years) and Residual ($44.1$ years) groups.

While the previous dimensions 1--3 were essential to divide the general public into the three categories defined in the pyramid model, responses to five additional questions in the NISTEP survey, corresponding to five additional predisposition dimensions, were examined to reveal more information about the types of people who fall into each category (Figure \ref{fig1}).
In these questions, respondents were asked to indicate the degree to which they agreed with each of the following five statements: \q{To reap the benefits of S\&T, we need to accept that a certain degree of risk accompanies S\&T issues} (dimension 4); \q{Humans' ability to control advanced or potentially dangerous S\&T is limited} (dimension 5); \q{S\&T has made our lives healthier, more convenient and more comfortable} (dimension 6); \q{The government should increase the budget for supporting R\&D of S\&T} (dimension 7); and \q{S\&T itself is harmless---it is only dangerous if it is used incorrectly or by the wrong people} (dimension 8).
All responses were recorded on a 4-point Likert scale from 4 (strongly agree) to 1 (strongly disagree).
These additional dimensions were useful to gain a better understanding about the level of S\&T awareness within each category of the general public.

We can see from Figure \ref{fig1} that those in the Attentive category indicated a strong belief in the value of S\&T, were favourably disposed to promoting S\&T policies and were willing to engage personally in the policymaking process.
In contrast, those in the Interested category appeared to have relatively passive attitudes towards S\&T and a strong preference that policymaking should be left to the experts.
The Interested group also appeared more comfortable than the Attentive group in accepting the outcomes of policymaking, presumably because these people rely on the judgements of experts.
Those in the Residual group appeared to have no strong preferences or specific opinions on any of the indicated dimensions, emphasising that, on average, this group is relatively indifferent or uninterested regarding S\&T policy.

\begin{figure}[!tp]
\centering
\vspace{-1em}
\noindent
\includegraphics[align=c, scale=1.0, vmargin=1mm]{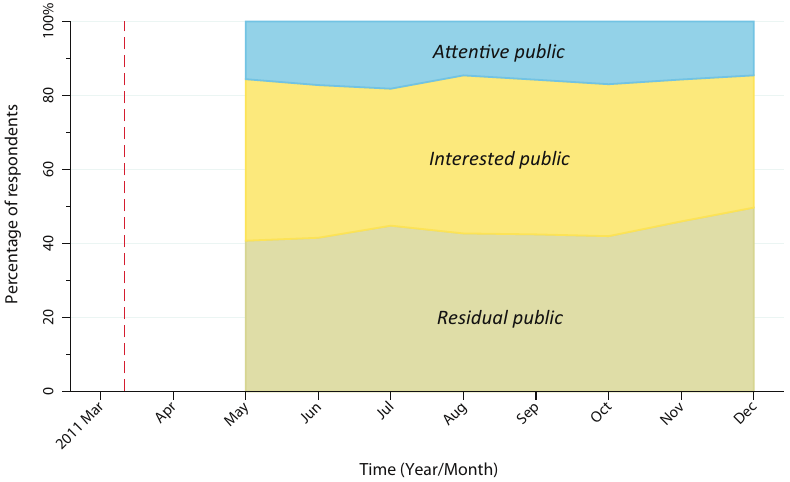}
%\vspace{6mm}
\caption{\textbf{Trend in the proportion of the general public falling into each group.}\\[0.2em]
\linespread{1.1}
{\footnotesize
The vertical dashed line indicates the date of the Great East Japan Earthquake (11 March 2011).
Surveys were conducted at the end of each month from April to November 2011, so they are recorded as \q{May} to \q{Dec} in the graph.}
}
\label{fig2}
\vspace*{1em}
\end{figure}

Finally, to check whether these categorisations for the general public are well defined over time, the aggregate sample population for each month of data was divided into the designated categories (Figure \ref{fig2}).
Although there were some variations over time, the proportion of people falling into each group was relatively stable, ranging from 15\% to 18\% for the Attentive group, 35\% to 44\% for the Interested group and 41\% to 50\% for the Residual group.
Fundamentally, the three categorisations for the general public remained well defined throughout the specified survey time period, in the sense that none of them expanded or shrank sufficiently as to challenge the three-category assumption.

%%%%%%%%%%
%%%%%%%%%%

%%%%%%%%%%%%%%%%%%%%%%%%%%%%%%%%%
%%%%%%%%%%%%%%%%%%%%%%%%%%%%%%%%%
\section{Drivers of public engagement in science policymaking}
%%%%%%%%%%%%%%%%%%%%%%%%%%%%%%%%%

Now that we have distinguished the general public into three groups according to their favoured form of engagement in S\&T policymaking, our next question is how specific socio-demographic factors or changes in some predisposition towards S\&T might drive individuals from one category to another.
To answer this question, we must first identify the key variables that are likely to determine whether people participate in the S\&T policymaking process.

%%%%%%%%%%%%%%%%%%%%%%%%%%%%%%%%%
\subsection{`Belief', `knowledge' and `trust'}
%%%%%%%%%%%%%%

Although identifying the salient exogenous variables in a comprehensive manner is difficult, my model, guided by an amalgam of conceptual, theoretical and methodological perspectives \cite{Ajzen:1991,Brossard:2007,Dijkstra:2012b,Evans:1995,Knight:2010,Pardo:2002,Priest:2001,RodutaRoberts:2013,Turcanu:2014,Wynne:1992}, hypothesises that three predictors, in addition to other socio-demographic variables, should be controlled for:
(1) \q{belief in S\&T-based society} (\q{belief} for short), indicating the extent to which people show a positive reaction to policies that promote and encourage the establishment of an S\&T-based society (see Table \ref{tab:keyQbelief} for summary statistics);
(2) \q{socio-scientific knowledge} (\q{knowledge}), describing people's knowledge about scientific issues, particularly those that are closely linked to their daily lives and society (Table \ref{tab:keyQknowledge});
and (3) \q{trust in S\&T experts} (\q{trust}), indicating the extent to which people deem S\&T experts' statements as trustworthy, with experts in this sense referring to both scientists and engineers (Table \ref{tab:keyQtrust}).
The modelling presented herein considered the fact that previous research has shown that trust in scientists can be influential in determining one’s acceptance of controversial S\&T policies, particularly in the absence of scientific knowledge \cite{Brewer:2011,Brossard:2007,Siegrist:2000a,Siegrist:2000b,Popkin:1994};
that is, in cases where people are not confident in their own knowledge, they are more likely to rely on S\&T experts whom they perceive as trustworthy.

Numerous studies have examined the relationships among \q{belief}, \q{knowledge} and \q{trust}.
However, in this study, no association or causation among these factors is presumed, for several reasons.

First, evidence linking these factors together is scant; there even appear to be conflicting dynamics between the different variables \cite{Achterberg:2014,Bauer:2000,Evans:1995}.
For instance, we are uncertain about the relationship between \q{belief} and \q{knowledge}, in that advanced knowledge has the potential to either increase or reduce a person's belief in S\&T.
In this case, it would not be appropriate to presume any association between these variables.

Second, reverse causation may also be a concern.
An advocate of S\&T policies might be more likely to acquire knowledge related to these issues.
However, the reverse scenario could also be true in that someone may become more supportive of S\&T policies as a result of knowledge.
In this case, it would be impossible to determine which of the two factors, \q{belief} or \q{knowledge}, was the cause and which was the effect.

Third, it would be more appropriate to regard any apparent association between the three predictors as being partly, if not completely, derived from the underlying common covariates.
From this perspective, we might expect the relationships between the three factors to be partly explained by their association with various socio-demographic factors.

Finally, there might be characteristics specific to the Japanese population that require consideration;
it will be necessary to think carefully about how to explain attitudes and behaviours in this context.
If this is the case, previous findings relevant to other countries or cultures cannot be assumed to be representative of the Japanese mindset with respect to S\&T issues.
Bearing in mind these considerations, this article does not assume any non-trivial relationships between the three key predictors used in the model.

%%%%%%%%%%%%%%%%%%%%%%%%%%%%%%%%%
\subsection{The model}
%%%%%%%%%%%%%%

\begin{figure}[!tp]
\centering
\vspace{-1em}
\noindent
\includegraphics[align=c, scale=1.05, vmargin=1mm]{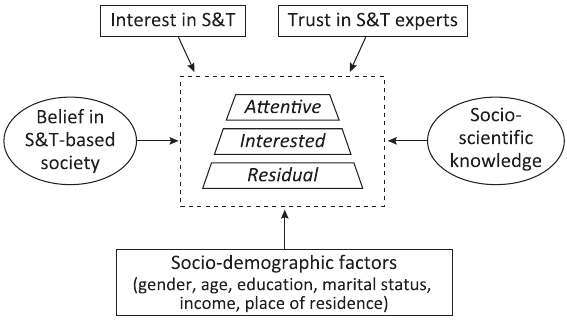}
%\vspace{6mm}
\caption{\textbf{Conceptual model for the formation of public attitudes towards S\&T policymaking.}\\[0.2em]
\linespread{1.1}
{\footnotesize
The observed responses for \q{interest} and \q{trust in S\&T experts} and the constructed latent variables \q{belief} and \q{knowledge} as well as socio-demographic variables together predict the likelihood that a person belongs to each of the three categories of the general public: Attentive, Interested and Residual.}
}
\label{fig3}
\vspace*{1em}
\end{figure}

The prior discussion leads to the development of the conceptual model used in this study, as depicted in Figure \ref{fig3} (see also Figure \ref{figS2}).
The variables corresponding to \q{trust in scientists} and \q{trust in engineers} were available directly from the NISTEP survey data and are labelled as \q{trust\_sci} and \q{trust\_eng}, respectively.
In contrast, the data did not contain questions that could act as a direct measure of belief and knowledge factors.
Therefore, these latent variables had to be constructed using other variables observed in the data via factor analyses.\endnote{%
Prior to conducting confirmatory factor analysis for each construct, the internal consistency of the scale was evaluated by the reliability coefficient (Cronbach's alpha), and the suitability of the data for factor analysis was assessed by the Kaiser--Meyer--Olkin measure of sampling adequacy and Bartlett's test of sphericity.
All measures and tests suggested that confirmatory factor analyses were appropriate for both constructs.
See Appendix \ref{app:FA} and Tables \ref{tab:FA:belief} and \ref{tab:FA:knowledge} for the technical details.}

The latent variable \q{belief} was constructed from the responses to four particular questions, in which respondents were asked to what extent they agreed with four different statements:
($S_{1}$) \q{It is important to develop researchers who are creative and have high ability levels}, ($S_{2}$) \q{It is important to actively support the universities' research and education programmes}, ($S_{3}$) \q{It is important that the government and public agencies conduct R\&D to solve social challenges} and ($S_{4}$) \q{It is important to promote the practical application of research results that bring innovation to society and economic activities}.
Respondents were able to answer using a 5-point Likert scale, from 5 (very much) to 1 (not at all).

%%%%%%%%%%
%%%%%%%%%%

The other latent variable, \q{knowledge}, was estimated using respondents' answers to four true--false statements, where the responses were coded as either 1 (correct) or 0 (incorrect or \q{don't know}).
These four statements tested respondents' general knowledge about radiation and nuclear energy.
Three of the statements were true: 
($Q_{2}$) \q{In 2009, Japan generated approximately 30\% of its domestic power from nuclear energy}, ($Q_{3}$) \q{Nuclear power produces no $\text{CO}_{2}$ emissions during electricity generating operations} and ($Q_{4}$) \q{People are exposed to natural radiation in their daily life}.
The remaining statement was false:
($Q_{1}$) \q{All radioactive substances are man-made or artificial}.
Since the survey data were collected in a period marked by widespread media attention to radiation and nuclear energy, it was consequently assumed that at least half the population would know which statements were true or false, which was indeed the case (see Table \ref{tab:keyQknowledge} for the percentage of correct answers to each question).
The collective use of these particular statements as proxies for people's socio-scientific knowledge is justifiable since radiation and nuclear energy issues were the S\&T issues most closely connected to the major event of concern, the Fukushima nuclear accident, during the survey period of April to November 2011.

The two latent variables \q{belief} and \q{knowledge}, constructed from $S_{1}$--$S_{4}$ and $Q_{1}$--$Q_{4}$, respectively, in addition to the three observed variables \q{interest}, \q{trust\_sci} and \q{trust\_eng} and socio-demographic variables together explain the likelihood that a person belongs to each of the three categories of the general public, that is, Attentive, Interested or Residual.
Given the unordered categorical nature of the criterion variable (i.e.\ category of the public), a multinomial logit regression model was constructed to test the hypothesis.\endnote{%
See Appendix \ref{app:mlogit} and Table \ref{tab:mlogit} for details of the multinomial logit regression analyses.}
The model also took into account the time-fixed effects, which were controlled for by including seven out of the eight survey months as dummy variables.

%%%%%%%%%%%%%%%%%%%%%%%%%%%%%%%%%
\subsection{Results}
%%%%%%%%%%%%%%

\begin{figure}[!tp]
\centering
\vspace{-1em}
\noindent
\includegraphics[align=c, scale=0.74, vmargin=1mm]{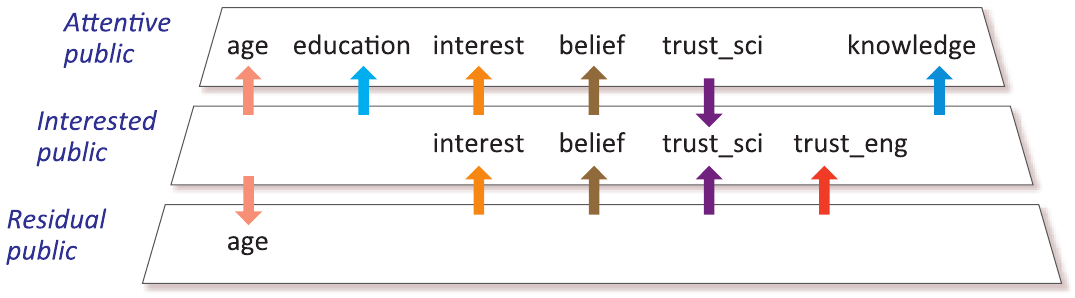}
%\vspace{6mm}
\caption{\textbf{Predicted effects: the drivers of switching among groups.}\\[0.2em]\linespread{1.1}
{\footnotesize
Each arrow indicates the direction in which an increase in the associated variable has a significant impact. For example, a person is more likely to be in the Attentive group than in the Interested group if the person is female, older, has a higher level of interest in S\&T, believes more strongly in an S\&T-based society, has greater socio-scientific knowledge or has less trust in scientists, holding other factors constant.}
}
\label{fig4}
\vspace*{1em}
\end{figure}

Figure \ref{fig4} schematically summarises the key findings from the regression analysis, which allows for an intuitive interpretation of the results as follows.\endnote{%
In my statements of the findings, the common conditional statement \q{holding other factors constant} is suppressed to save space.}
First, older people are more likely to belong to either the Attentive or the Residual group than to the Interested group ($p<.001$).
In other words, ageing has the effect of pushing individuals from the Interested category to either of the two outer categories of the general public in the pyramid model (albeit not necessarily a causal effect;
the same caveat applies to the following observations).
We could call this \q{outward pressure}.

Second, those with higher levels of interest or belief in S\&T are more likely than those with lower levels to gravitate towards groups where there is stronger engagement with S\&T:
Attentive relative to Interested ($p<.001$ for \q{interest} and $p=.002$ for \q{belief}) and Interested relative to Residual ($p<.001$ for both \q{interest} and \q{belief}).
Namely, they are more likely to be pushed upwards in the pyramid.

Third, interestingly, trust in scientists appears to lead to \q{inward pressure};
that is, it drives individuals in the Attentive and Residual groups towards the middle layer, Interested ($p<.001$).
Note that trust in scientists does not appear to play a role in moving individuals from the Interested to the Attentive group; conversely, individuals in the Attentive group are more likely to move down to the Interested group if they have higher trust levels.
In contrast, trust in engineers appears to have only the effect of pushing individuals in the Residual group up to the Interested group ($p=.002$).

Finally, educational attainment and socio-scientific knowledge are important factors only in driving individuals from the Interested to the Attentive group ($p=.001$ for \q{education} and $p<.001$ for \q{knowledge});
these do not push people from the Residual into the Interested group ($p=.154$ for \q{education} and $p=.695$ for \q{knowledge}).

By contrast, \q{gender}, \q{marital status} and \q{income} are never significant factors in explaining the dynamics of movement either between the Attentive group and the Interested group or between the Interested group and the Residual group.
The result about the gender effect is particularly noteworthy as it is contrary to the common expectation that men are more interested in S\&T and more willing to engage with S\&T policymaking;
indeed, we previously observed that 67\% of those in the Attentive group were male.
However, once other independent variables are controlled, we can see the direct effect of gender, which is insignificant within the current modelling framework.

%%%%%%%%%%
%%%%%%%%%%

%%%%%%%%%%%%%%%%%%%%%%%%%%%%%%%%%
\subsection{Implications for Japan's post-disaster reactions}
%%%%%%%%%%%%%%

The aforementioned findings can offer insights into how the 2011 earthquake and nuclear disaster affected public attitudes towards S\&T policymaking in Japan.
First, it is notable that trust in S\&T experts dropped dramatically just after the occurrence of the earthquake \citex{\textcolor{black}{~see also Figure \ref{figS1}}}{MEXT:2012}.
This fact, together with the regression results, suggests that during this short time period the middle layer of the general public, those identified as being in the Interested category, must have been subject to an \q{outward pressure} in the three-layered pyramid, towards either the Attentive or the Residual category.
Second, during the same period, the public's level of interest in S\&T sharply increased (Figure \ref{figS3}).
On the basis of the regression results, this should have resulted in an overall \q{upward pressure}, driving the public towards the apex of the policymaking pyramid.
Finally, during the same period, the public are likely to have increased their levels of socio-scientific knowledge, namely, their understanding of disaster prevention and reduction and their knowledge of nuclear power and radioactivity issues.
This is likely due to the significant media coverage about the earthquake, tsunami and nuclear power plant accident.
The regression results then imply that those in the Interested category must have been subject to another \q{upward pressure} towards the Attentive category.

As a result of these anticipated effects, it is plausible that following the earthquake, a certain portion of people who used to be in the Interested category would have consequently moved to the Attentive category with an increased level of interest in S\&T and/or socio-scientific knowledge as well as higher levels of distrust for experts.
Conversely, others who simply lost trust in experts without increasing their level of interest and/or knowledge would have moved to the Residual category.
Indeed, this appears to explain the trend seen in Figure \ref{fig2} for the three months immediately after the earthquake (from \q{May} to \q{Jul}), where the proportion of the Interested public monotonously shrank (from 43.7\% to 37.0\%; $p=.004$), while the other two outer categories expanded.
Recall that no information about people's level of trust or socio-scientific knowledge was used to define the categories of the general public.
This means that the consistency between the preceding argument, using the regression analysis, and the descriptive observations presented in Figure \ref{fig2} is non-trivial and thus may provide additional support for the validity of the regression modelling and the resulting interpretations.

It can be further observed that for the last three months shown in Figure \ref{fig2} (from \q{Oct} to \q{Dec}), the proportion of people within the Residual category monotonously increased (from 42.0\% to 49.7\%; $p=.001$), while the proportions of the other two categories decreased.
The regression results also help to explain this trend: it may be that people gradually lost interest in S\&T-related news pertinent to the disaster and/or lost belief in an S\&T-based society.

These insights into the heterogeneous behavioural responses of the general public could help S\&T stakeholders including policymakers and journalists, who have been engaged with the post-disaster rebuilding of Japan's science communication system, to design and implement a more effective and accountable S\&T policymaking process.

%%%%%%%%%%%%%%%%%%%%%%%%%%%%%%%%%
%%%%%%%%%%%%%%%%%%%%%%%%%%%%%%%%%
\section{Conclusion and discussion}
%%%%%%%%%%%%%%%%%%%%%%%%%%%%%%%%%

The key contributions of this study are twofold.
First, it reveals that members of the general public can meaningfully be categorised as Attentive, Interested or Residual based on their predispositions towards S\&T and other socio-demographic characteristics; 
second, it identifies several key drivers influencing the shift of individuals from one category of the general public to another. 
On the basis of these findings, it furthermore provides some possible explanations regarding the impact of Japan's 2011 disasters on individuals in each category.

Finally, this article closes with a discussion on policy implications of the findings, which subsequent studies might help to refine, as well as the validity and limitations of this study.

%%%%%%%%%%%%%%%%%%%%%%%%%%%%%%%%%
\subsection{Towards evidence-based science communication policymaking}
%%%%%%%%%%%%%%

Insights drawn from this study provide a foundation for understanding the relationships between the mediators of S\&T and the general public.
These insights would be useful, for example, to clarify and re-define the roles of distinct types of science communicators who interact organically with all S\&T stakeholders at different levels of the policymaking pyramid. 
In this regard, the findings of this study could be helpful in designing and implementing an effective approach to achieving particular policy goals by focusing on the distinct roles of science communicators.

Suppose, for instance, that the government professes to advocate an S\&T-intensive society and therefore aims to increase national public awareness of S\&T policy issues. 
In terms of the pyramid model, this activity would have to involve encouraging people currently in the Residual group to move into the Interested category. 
Based on the regression results, policy interventions that seek simply to educate or enlighten people by increasing their levels of socio-scientific knowledge are not likely to be effective in this regard since \q{knowledge} did not have a significant effect in causing a person to move from the Residual to the Interested group. 
Rather, policies that enhance people's support for S\&T appear more appropriate for this purpose, particularly considering that \q{belief} was significant in producing movement both from Residual to Interested and from Interested to Attentive. 
In this case, science communicators' activities designed to stimulate public curiosity, such as engaging with science caf\'{e}s and festivals or developing and supporting exhibitions at museums or within local communities, should effectively increase public support for S\&T through generating scientific enthusiasm.

In another scenario, suppose that policymakers established a priority of encouraging public participation in S\&T policymaking dialogue. 
In this case, the predictions from the regression model suggest that policies aimed at increasing levels of socio-scientific knowledge should be effective in stimulating people to shift from the Interested group to the Attentive category. 
Policies that increase public interest in S\&T issues would also be effective both directly, by inducing people to move from the Interested to the Attentive group, and indirectly by encouraging movement from the Residual category to the Interested category. 
In the latter case, further movement into the Attentive category is also possible if the driving stimuli are seen as sufficiently relevant. 
In this scenario, policy-oriented activities by science communicators could be most useful. 
Communicators could report on the impact of government S\&T policies from various perspectives, such as academic import, cost-effectiveness, balance in resource allocation and comparative advantages or competitiveness. 
Such activities will allow those in the Interested group to access a broad range of information about policy alternatives and to acquire socio-scientific knowledge that is closely linked to their daily lives and society.

To be sure, the scenarios and recommendations described here would need considerable refinement in actual practice; 
however, more detailed investigation along these lines should yield deeper insight into the heterogeneous nature of the general public's predispositions and responses to S\&T issues, resulting in useful guidance for the roles of science communicators. 
This information will further enable us to explore, more fully understand and more effectively exploit the real-world science communication ecosystem and its dynamics.

%%%%%%%%%%%%%%%%%%%%%%%%%%%%%%%%%
\subsection{Validity and limitations}
%%%%%%%%%%%%%%

As with any quantitative social science research, this study faces limitations that may have impacted the general validity of the findings. 
First, the way in which the three categories of the general public were identified in the cluster analysis would need further refinement. 
This article employed three predisposition variables---\q{interest in S\&T}, \q{participation of the general public} and \q{delegation to experts}---to define the clusters. 
However, other observed variables, which are directly available in the NISTEP dataset, and possible latent variables, which can be constructed from the observed ones, would have also been useful to improve the cluster analysis so that one could grasp the distribution of people across the designed categories in a more accurate and reliable manner.

Second, the regression model, while relatively comprehensive, is unlikely to capture every salient predictor of the public's attitudes towards S\&T policymaking. 
For example, predispositions relating to people's trust in government institutions and actors or political leaders \citec{\textcolor{black}{as studied by, for example,}}{}{Brewer:2011,Liu:2009,Siegrist:2000a,Vainio:2013}, their environmental beliefs \cite{Achterberg:2014,Achterberg:2010}, their political or economic ideology \cite{Becker:2010,Brewer:2011}, their religious beliefs \cite{Becker:2010,Brossard:2009,Gaskell:2005}, their perceptions about the fairness of decision-making procedures \cite{Besley:2010} or their \q{political efficacy} or \q{belief in public efficacy} \cite{Knight:2010} were not considered by the NISTEP questionnaire; 
as such, these were not discussed in this article. 
These omitted variables may have affected the regression results as they may be associated with both the criterion variable (i.e.\ the category of the general public) and some of the explanatory variables.

In addition, a closer examination of the validity of each explanatory variable used in the regression analysis would be necessary to render the model more capable of capturing causal effects. 
For example, in the analysis within this article, a person's knowledge about radiation and nuclear energy was used as a proxy for \q{socio-scientific knowledge}, one of the key hypothesised explana tory variables. 
Although there is some justification for this approach, as explained above, a more comprehensive latent construct such as \q{scientific literacy} \cite{Miller:1998,Miller:2004} or \q{contextualised (socio-)scientific knowledge} \cite{Sturgis:2004} might be more appropriate.

Another source of bias would have arisen had the conceptual model of Figure \ref{fig3} been seen to have an incomplete path-structure. 
For example, it might have been more accurate to assume that certain predispositions, such as \q{belief}, \q{knowledge} and \q{trust}, were partly explained by individual socio-demographic variables, rather than acting as exogenous variables, as implied in previous studies \citec{\textcolor{black}{e.g.}\/}{}{Bauer:2000,Brewer:2011,Brewer:2013,Brossard:2007,Miller:2007,Priest:2001,RodutaRoberts:2013}. 
In other words, these socio-demographic variables might have had both direct and indirect effects, with the indirect effects being mediated by some predisposition towards S\&T, which then directly affects public attitudes towards S\&T policymaking.\endnote{%
A more refined approach to deal with this issue may be to use structural equation modelling, which allows estimating and testing the entire model simultaneously, including both the direct and indirect effects of socio-demographic variables and relevant predispositions.}
It could also be the case that some interaction effects \cite{Achterberg:2014,Allum:2008,Brossard:2009,Liu:2009,vonRoten:2004} play a role in explaining the public attitudes.

Moreover, the method of participant recruitment may have limited the validity of the findings since it may be difficult to make inferences about the full population from these data. 
Since the NISTEP data used in this study were based on an Internet survey, the findings of the study hold only within the specific characteristics of the sample. 
In fact, on average, this sample has higher educational attainment than the overall Japanese population; 45\% of the respondents are regarded as having higher education, whereas the corresponding percentage in the latest national census was 20\% \cite{kokusei:2010}. 
In such a case, according to previous studies by \citek{NISTEP-report:2008,NISTEP:2010}, the sample of respondents is likely to be skewed towards more knowledgeable people who are also more likely to be supportive of S\&T. 
In addition, the same studies suggest that Japanese people tend to show more interest and positive attitudes towards S\&T in Internet surveys than in face-to-face interviews. 
Due to these external validity issues, one should exercise caution in generalising the results observed in these samples to the entire population of Japan, let alone to other countries or cultures and/or different times.

It is hoped that future studies will reveal more about the mechanisms of public attitude formation by overcoming some of the above-mentioned limitations and caveats through further conceptual, theoretical and methodological improvements. 
Such continued efforts could contribute towards establishing a more effective public engagement platform that would in turn induce more accountable, evidence-based policymaking, thereby shedding new light on the relationship between S\&T and society.

%\clearpage
%\nolinenumbers
\titleformat*{\section}{\large\bfseries}

\vspace{1.5em}
%%%%%%%%%%%%%%%%%%%%%%
%          ACKNOWLEDGEMENTS
%%%%%%%%%%%%%%%%%%%%%%
\noindent
{\large \textbf{Acknowledgements}}\\[0.5em]
{\small The author would like to thank M.W.~Bauer, P.~Dunleavy, D.~Rinnert, E.~Spencer and the two anonymous reviewers for their suggestions and comments. 
This article is dedicated to all those who have committed, and who continue to commit, to the rebuilding of Japan following the Great East Japan Earthquake on 11 March 2011 and the subsequent Fukushima nuclear power plant accident.
All content is the responsibility of the author and does not necessarily reflect the views of the MEXT or NISTEP.
%The views and conclusions contained herein are those of the author and should not be interpreted as necessarily representing the official policies or endorsements, either expressed or implied, of any of the organisations to which the author is affiliated.
}

\vspace{1.5em}
%%%%%%%%%%%%%%%%%%%%%%
%          FUNDING INFORMATION
%%%%%%%%%%%%%%%%%%%%%%
\noindent
{\large \textbf{Funding}}\\[0.5em]
{\small The author received no financial support for the research, authorship and/or publication of this article.}

\vspace{-0.5em}
%%%%%%%%%%%%%%%%%%%%%%
%          ENDNOTES
%%%%%%%%%%%%%%%%%%%%%%
\renewcommand{\enotesize}{\small}
\theendnotes
\addcontentsline{toc}{section}{Notes}

%%%%%%%%%%%%%%%%%%%%%%
%          REFERENCES
%%%%%%%%%%%%%%%%%%%%%%
\bibliographystyle{apalike-imp}
\addcontentsline{toc}{section}{References}

\setlength{\bibsep}{0\baselineskip plus 0.2\baselineskip}
\renewcommand*{\bibfont}{\footnotesize}

\vspace{3.0em}
%%%%%%%%%%%%%%%%%%%%%%
%          AUTHOR BIOGRAPHY
%%%%%%%%%%%%%%%%%%%%%%
\noindent
{\large \textbf{Author biography}}\\
\textsf{Keisuke Okamura} (PhD) is a government official at the Science and Technology Policy Bureau of the Ministry of Education, Culture, Sports, Science and Technology (MEXT), Japan. 
From 2011 to 2012, he served as a member of the Government's Nuclear Emergency Response Team in response to the Fukushima nuclear accident.

%%%%%%%%%%%%%%%%%%%%%%%%%%%%%%%%%%%%%%%%%%%%%%%%%%%%%%%%%%%%%%%%%%%%
%\clearpage
%\fancyhead[LE,RO]{\textcolor{orange}{}}
%
%\begin{center}
%\textcolor{VioletRed}{\textsf{[THIS PAGE INTENTIONALLY LEFT BLANK]}}
%\end{center}
%%%%%%%%%%%%%%%%%%%%%%%%%%%%%%%%%%%%%%%%%%%%%%%%%%%%%%%%%%%%%%%%%%%%

\clearpage
\newpage
%%%%%%%%%%%%%%%%%%%%%%%%%%%%%%%%%%%%%%%%%%%%
%%%%%%%%%%%%%%%%%%%%%%%%%%%%%%%%%%%%%%%%%%%%
%          SUPPLEMENTARY MATERIALS
%%%%%%%%%%%%%%%%%%%%%%%%%%%%%%%%%%%%%%%%%%%%
%%%%%%%%%%%%%%%%%%%%%%%%%%%%%%%%%%%%%%%%%%%%

%\renewcommand{\thesection}{\Alph{section}}
\renewcommand{\thesubsection}{Appendix\,~\Alph{subsection}}
\renewcommand{\thesubsubsection}{\Alph{subsection}.\,\arabic{subsubsection}}

\pagestyle{fancy}
\fancyhead[LE,RO]{\textcolor{orange}{\footnotesize{\textsf{SUPPLEMENTARY MATERIALS}}}}
\fancyhead[RE,LO]{}
\fancyfoot[RE,LO]{\color[rgb]{0.04, 0.73, 0.71}{}}
\fancyfoot[LE,RO]{\scriptsize{\textbf{\textsf{\thepage}}}}
\fancyfoot[C]{}

\addcontentsline{toc}{section}{Supplementary Materials}

%%%%%%%%%%%%%%%%%%%%%%
\quad
\vspace{-0.5cm}
\begin{center}
\fontsize{15pt}{16pt}\selectfont\bfseries
%%%%%%%%%%%%%%%%%%%%%%
%%%%%%%%%%%%%%%%%%%%%%
Supplementary Materials
%%%%%%%%%%%%%%%%%%%%%%
%%%%%%%%%%%%%%%%%%%%%%
\end{center}

\vspace{1.0cm}

%\clearpage
%%%%%%%%%%%%%%%%%%%%%%%%%%%%%
%\renewcommand{\figurename}{Suppl.~Figure}
%\renewcommand{\tablename}{Suppl.~Table}
\renewcommand{\thefigure}{S\arabic{figure}}
\renewcommand{\thetable}{S\arabic{table}}
\renewcommand{\theequation}{S\arabic{equation}}
\setcounter{section}{0}
\setcounter{subsection}{0}
\setcounter{figure}{0}
\setcounter{table}{0}
\setcounter{equation}{0}

\renewcommand{\headrule}{\color{orange}\oldheadrule}

\vspace{-0.5cm}
%%%%%%%%%%%%%%%%%%%%%%%%%%
\begin{mdframed}[linecolor=gray]
\begin{tabular}{l}
\begin{minipage}{0.98\hsize}
\setstretch{1.0}
\textcolor{gray}{\footnotesize\textsf{This supplementary document provides the methodological details of the factor analyses and the multinomial logit regression analyses discussed in the paper \textit{\q{\mtitle}} (DOI: \href{https://journals.sagepub.com/doi/10.1177/0963662515605420}{10.1177/0963662515605420}) by Keisuke Okamura published in \textit{Public Understanding of Science}, {25}(4):465--479, 2016. 
It also contains Supplementary Figures \ref{figS1}--\ref{figS3}, Supplementary Tables \ref{tab:month-region}--\ref{tab:mlogit} and References.}}
\end{minipage}
\end{tabular}
\end{mdframed}
%%%%%%%%%%%%%%%%%%%%%%%%%%

\vspace{0.2cm}

%%%%%%%%%%%%%%%%%%%%%%
%%%%%%%%%%%%%%%%%%%%%%
\subsection[{\hspace{3.5eM}} Details of data analyses]{Details of data analyses\label{app:method}}

Throughout the paper, all data were analysed with STATA/IC 13.0 software (StataCorp LP, Texas, USA). 
Frequency tables, cross-tabulations, cluster analysis, factor analyses, and multinomial logit regression analyses were conducted. 
The basic significance level for tests was 0.05. 
In all statistical analyses, it was ensured that the sampling weights were properly accounted for, thereby compensating for the sample design that may have over or underrepresented various segments within the population.

%%%%%%%%%%%%
\subsubsection[{\hspace{0.5eM}} Factor analyses]{Factor analyses\label{app:FA}}

Prior to conducting confirmatory factor analysis for the latent variable \q{belief}, the internal consistency of the responses to the four constituent items $S_{1}$--$S_{4}$ (see Table \ref{tab:keyQbelief} for summary statistics) was evaluated by Cronbach's alpha reliability coefficient. 
The resulting alpha coefficient for $S_{1}$--$S_{4}$ was 0.89, which inferred that there was strong consistency in respondents' answers across the questions (Hair et al., 1998). 
In addition, Bartlett's test of sphericity was significant (${\chi}^{2}=14086.2$ at $p<0.001$) and the Kaiser--Meyer--Olkin (KMO) measure of sampling adequacy was $0.80$, which showed an adequate fit \cite{Hair:1998}. 
Thus, these tests suggested that there were sufficient inter-item correlations within the data for performing factor analysis. 
The predicted score for \q{belief} was computed from the loadings of the constituent measurements. 
These results are summarised in Table \ref{tab:FA:belief}.

The same method was used to measure the other latent variable \q{knowledge}. 
This time the check for internal consistency of the responses to the four constituent items $Q_{1}$--$Q_{4}$ (see Table S5 for summary statistics) was completed using the Kuder--Richardson formula 20 (KR20), a special case of Cronbach's alpha in which the items are binary. 
This indicator of internal consistency can be thought of as a measure of how consistent the responses to the test-style questions are in providing information about a respondent’s level of socio-scientific knowledge, with respect to the content assessed by the statements. 
The resulting KR20 coefficient for $Q_{1}$--$Q_{4}$ was 0.63, which was not particularly high but remained at an acceptable level for internal consistency validity (Hair et al., 1998). 
Again, Bartlett's test of sphericity (${\chi}^{2}=2756.4$ at $p<0.001$) and the KMO measure of sampling adequacy ($\text{KMO}=0.72$) were examined, both of which suggested that the use of factor analysis was appropriate (Hair et al., 1998). 
These results are summarised in Table \ref{tab:FA:knowledge}.

%%%%%%%%%%%%
\subsubsection[{\hspace{0.5eM}} Multinomial logit regression analyses]{Multinomial logit regression analyses\label{app:mlogit}}

The conceptual path diagram presented in Figure \ref{fig3} can be seen in terms of defined variables in the regression model; see Figure \ref{figS2}. 
The Interested category was used as the base category within the three-outcome multinomial logit model. 
This choice of the base outcome specifies that the coefficients associated with the other outcomes are to be measured relative to that base. 
Specifically, the probability of each outcome is to be given by
\begin{subequations}
\begin{alignat}{3}
&{\rm Pr}(\text{Attentive}\!\mid\!\bmt{x})&{}={\rm Pr}({\tt group}=1\!\mid\!\bmt{x})&=\exp(\bmt{x}\bmt{\beta}_{1})\big/D\,,\label{eq:1a}\\
&{\rm Pr}(\text{Interested}\!\mid\!\bmt{x})&={}{\rm Pr}({\tt group}=2\!\mid\!\bmt{x})&=1/D\,,\\
&{\rm Pr}(\text{Residual}\!\mid\!\bmt{x})&{}={\rm Pr}({\tt group}=3\!\mid\!\bmt{x})&=\exp(\bmt{x}\bmt{\beta}_{3})\big/D\,,\label{eq:1c}
\end{alignat}
\end{subequations}
where $\bmt{x}=(x_{1},\dots,x_{k})$ is a $1\times k$ vector of independent variables, $\bmt{\beta}_{m}=(\beta_{m1},\dots,\beta_{mk})'$ is a $k\times 1$ vector of regression coefficients ($m=1,\,3$) which are the unknown population parameters to be estimated, and $D=1+\exp(\bmt{x}\bmt{\beta}_{1})+\exp(\bmt{x}\bmt{\beta}_{3})$ is a normalisation factor such that the probabilities sum to 1.
The logarithm of the odds (the ratio of the probability of choosing the Attentive or Residual category over the probability of choosing the base category, Interested) follows from the relationships (\ref{eq:1a})--(\ref{eq:1c}):
\begin{align}
\ln\ko{\f{\mathrm{Pr}(\text{Attentive})}{\mathrm{Pr}(\text{Interested})}}=\bmt{x}\bmt{\beta}_{1}\quad \text{and}\quad 
\ln\ko{\f{\mathrm{Pr}(\text{Residual})}{\mathrm{Pr}(\text{Interested})}}=\bmt{x}\bmt{\beta}_{3}\,,
\end{align}
namely, as a linear combination of the independent variables.
Note that the \q{Independence from Irrelevant Alternatives (IIA)} property of the multinomial logit model is assumed to hold; the alternative outcomes (i.e.\ \q{Attentive}, \q{Interested} and \q{Residual}) are considered dissimilar and therefore not substitutes for one another. 
In fact, the results of the Small--Hsiao test ($p=0.26$ for \q{Attentive}, $p=0.61$ for \q{Interested} and $p=0.66$ for \q{Residual}) provided evidence that the IIA assumption was not violated.

For comparability, four different models corresponding to different specifications of independent variables were explored and analysed. They are labelled as Model 1 to Model 4 with the following sets of independent variables defined respectively for each model:
\begin{subequations}
\begin{align}
\bmt{x}^{(1)}&=
({\tt female} ,\, {\tt age} ,\, {\tt educ} ,\, {\tt married}  ,\, {\tt income} ,\, {\tt Region},\, {\tt YearMonth},\, {\tt interest})\,,\label{eq:model1}\\
\bmt{x}^{(2)}&=
(\bmt{x}^{(1)} ,\, {\tt trust\_sci} ,\, {\tt trust\_eng})\,,\label{eq:model2}\\
\bmt{x}^{(3)}&=
(\bmt{x}^{(2)} ,\, {\tt knowledge})\,,\label{eq:model3}\\
\bmt{x}^{(4)}&=
(\bmt{x}^{(3)} ,\, {\tt belief})\,.\label{eq:model4}
\end{align}
\end{subequations}
In Model 1, socio-demographic variables, such as gender, age, educational attainment, marital status, income level, and place of residence (\q{region}) were used as independent variables, as well as \q{interest in S\&T}. 
As for the variable \q{Region}, which takes on one of the seven regions of Japan (\q{Hokkaido}, \q{Tohoku}, \q{Kanto}, \q{Chubu}, \q{Kinki}, \q{Chu-Shikoku} and \q{Kyushu}), \q{Tohoku} was used as the base category, while the other regions were used as controls (i.e.\ six dummy variables). 
The model also included time-fixed effects (\q{YearMonth}), which were controlled for by including seven out of the eight survey months (from April to November 2011) as dummy variables. 
In Model 2, variables associated with \q{trust in S\&T experts} were included as additional independent variables. 
Similarly, Models 3 and 4 each represented the prior model, plus the addition of a new independent variable; \q{socio-scientific knowledge} and \q{belief in S\&T-based society} were added respectively.

The results of the multinomial logit regression analyses are summarised in Table \ref{tab:mlogit}, where the outcome category is Attentive in part (a) and Residual in part (b). 
In demonstrating how to interpret the results, let us look at the first regression coefficient, \q{$-0.200^{*}$}, shown in part (a), \q{female} in Model 1. 
This represents the logarithm of the \q{odds ratio}, where the \q{odds} are defined as before, and the \q{ratio} reflects the situation when the dummy variable \q{female} takes the value 1, relative to when it takes 0, holding other variables constant. In mathematical terms, this implies that:
\begin{equation}
\left.\ko{\f{\mathrm{Pr}(\text{Attentive})}{\mathrm{Pr}(\text{Interested})}}_{\mathrm{female}}\right/
\ko{\f{\mathrm{Pr}(\text{Attentive})}{\mathrm{Pr}(\text{Interested})}}_{\mathrm{male}}
=e^{-2.00}\approx 0.819\,.
\end{equation}

\noindent
This result suggests that, based on the multinomial logit model, the probability ratio of being in the Attentive category, relative to the Interested category, is approximately 18\% lower for women than for men, holding all other factors constant. 
It is also indicated (by the asterisk) that the association is statistically significant at the 5\% level.

Based on the pseudo-$R^{2}$ for each model (see the bottom row of part (b)), Model 4 was the preferred model in terms of goodness-of-fit, having the highest pseudo-$R^{2}$ value. 
Subsequently, therefore, this was the model considered in more detail in the present study. 
By examining the sign and statistical significance of the coefficients, the key findings from part (a) of Table \ref{tab:mlogit} can be summarised as follows:
\begin{itemize}
\setlength\itemsep{-0.4em}
\item Older people are more likely to belong to the Attentive group than to the Interested group.
\item Those with higher levels of \q{interest}, \q{education}, \q{knowledge} and \q{belief} are more likely to belong to the Attentive group, relative to the Interested group, than those with lower levels.
\item Those with higher levels of \q{trust in scientists} are less likely to belong to the Attentive group, relative to the Interested group, than those with lower levels of trust.
\end{itemize}
Similarly, the key findings from part (b) are summarised as follows:
\begin{itemize}
\setlength\itemsep{-0.4em}
\item Older people are more likely to belong to the Residual group than to the Interested group.
\item Those with higher levels of \q{interest}, \q{belief} and \q{trust in S\&T experts (both scientists and engineers)}, are less likely to belong to the Residual group, relative to the Interested group, than those with lower levels.
These findings can be cast into the schematic diagram of Figure \ref{fig4}.
\end{itemize}

%%%%%%%%%%%%%%%%%%%%%%%%%%%%
%%%%%% SUPPLEMENTARY REFERENCES %%%%%%%
%%%%%%%%%%%%%%%%%%%%%%%%%%%%

%%%%%%%%
\renewcommand\refname{\fontsize{11}{15}\selectfont References for Supplementary Materials}

\setlength{\bibsep}{0\baselineskip plus 0.2\baselineskip}
\renewcommand*{\bibfont}{\footnotesize}
\renewcommand{\baselinestretch}{1.2}

%%%%%%%%%%%%%%%%%%%%%%%%%%%%
%%%%%% SUPPLEMENTARY FIGURES %%%%%%%
%%%%%%%%%%%%%%%%%%%%%%%%%%%%
\addcontentsline{toc}{subsection}{Supplementary Figures}

%+++++++++++++++++++++++++++++++++++++++++++++++++++++++++++++++++++++++++++
\afterpage{\clearpage%
%%%%%%%%%%
\begin{figure}[!tp]
\centering
%\vspace{-0.5cm}
\noindent
\includegraphics[align=c, scale=1.15, vmargin=1mm]{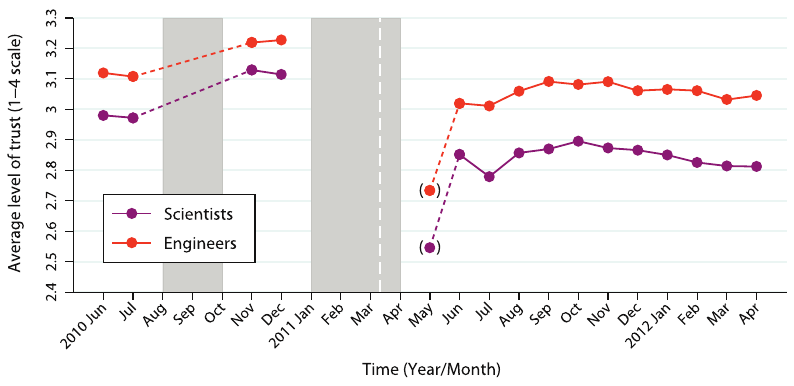}
%\vspace{6mm}
\caption{\textbf{The trend in the average level of public trust in S\&T experts.}\\[0.2em]
\linespread{1.1}{\footnotesize
The vertical dashed line indicates the date of the Great East Japan Earthquake (11 March 2011). 
No corresponding survey questions were asked for the months within the areas shaded grey. 
The question statement used was: \q{Do you find scientists' statements trustworthy?} Respondents were asked to select from one of the following five response categories: from 4 (very trustworthy) to 1 (very untrustworthy) and \q{don't know}. 
In creating the graph, the \q{don't know} responses were dropped before computing the average scores for each month.\\[1.5em]
\textit{Note:}
The April 2011 survey, recorded as \q{2011 May} within the graph, used a different questionnaire format from other months, asking respondents to indicate the degree to which they agree with the statement: \q{scientists/engineers are trustworthy}. 
Respondents were provided with the following five response categories: from 4 (strongly agree) to 1 (strongly disagree) and \q{don't know}. 
Therefore, it is not appropriate to directly compare this with the results taken for other months.\\[1em]
\textit{Source:}
Created by the author, based on the \citek{NISTEP-data:2012app} data.
}}
\label{figS1}
\vspace{5mm}
\end{figure}
%%%%%%%%%%
\quad\\
\vfill
}

%+++++++++++++++++++++++++++++++++++++++++++++++++++++++++++++++++++++++++++
\afterpage{\clearpage%
%%%%%%%%%%
\begin{figure}[!tp]
\centering
%\vspace{-0.5cm}
\noindent
\includegraphics[align=c, scale=0.95, vmargin=1mm]{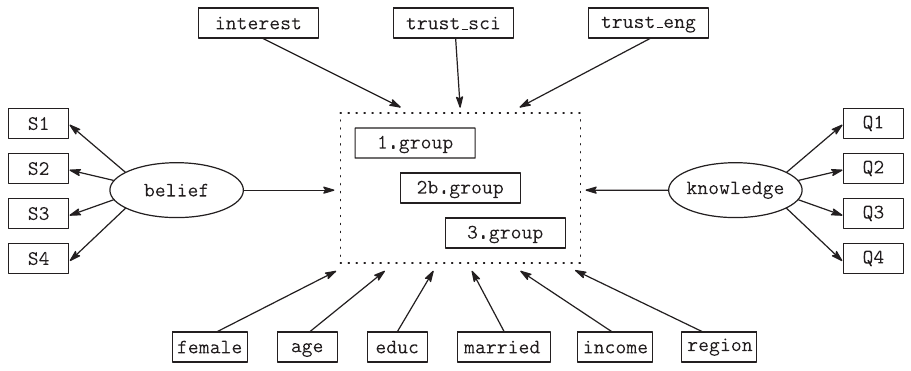}
%\vspace{6mm}
\caption{\textbf{A multinomial logit regression model for formation of public attitudes towards S\&T policymaking.}\\[0.2em]
\linespread{1.2}{\footnotesize
The observed variables \q{\texttt{interest}}, \q{\texttt{trust\_sci}} and \q{\texttt{trust\_eng}}, and socio-demographic variables \q{\texttt{female}}, \q{\texttt{age}}, \q{\texttt{educ}}, \q{\texttt{married}}, \q{\texttt{income}} and \q{\texttt{region}}, as well as the predicted latent variables \q{\texttt{belief}} and \q{\texttt{knowledge}}, together explain the likelihood that a person belongs to each category of the general public: Attentive (\q{\texttt{1.group}}), Interested (\q{\texttt{2b.group}}) or Residual (\q{\texttt{3.group}}). 
The \q{\texttt{b}} in \q{\texttt{2b.group}} indicates that this group, the Interested public, is used as the base category.}}
\label{figS2}
\vspace{5mm}
\end{figure}
%%%%%%%%%%
\quad\\
\vfill
}

%+++++++++++++++++++++++++++++++++++++++++++++++++++++++++++++++++++++++++++
\afterpage{\clearpage%
%%%%%%%%%%
\begin{figure}[!tp]
\centering
%\vspace{-0.5cm}
\noindent
\includegraphics[align=c, scale=1.15, vmargin=1mm]{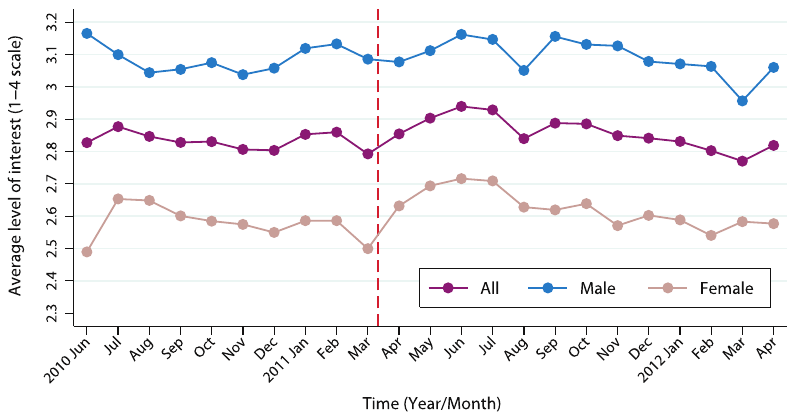}
%\vspace{6mm}
\caption{\textbf{Trend in the average level of public interest in S\&T.}\\[0.2em]
\linespread{1.2}\footnotesize
The vertical dashed line indicates the date of the Great East Japan Earthquake (11 March 2011). 
The question asked the respondents to indicate the degree to which they are interested in S\&T news and issues, allowing them to select from one of the following five response categories: from 4 (very interested) to 1 (not interested at all) and \q{don't know}. 
In creating the graph, the \q{don't know} responses were dropped before computing the average scores for each month.\\[1em]
\textit{Source:}
Created by the author, based on the \citek{NISTEP-data:2012app} data.
}
\label{figS3}
\vspace{5mm}
\end{figure}
%%%%%%%%%%
\quad\\
\vfill
}

\newpage
%%%%%%%%%%%%%%%%%%%%%%%%%%%%
%%%%%% SUPPLEMENTARY TABLES %%%%%%%
%%%%%%%%%%%%%%%%%%%%%%%%%%%%
\addcontentsline{toc}{subsection}{Supplementary Tables}

{\clearpage%
%%==================================================
\vspace*{-3em}
\centering
\begin{threeparttable}
\renewcommand{\baselinestretch}{1.1}
\caption{\textbf{Number of respondents by month and region.}}
\label{tab:month-region}
{\footnotesize
\begin{tabularx}{400pt}{c *9{>{\Centering}X}}\toprule\\[-1.5em]
\makecell{Month \\[-0.2em] (2011)\\[-1.3em]} & Kyushu & Chu-Shikoku & Kinki & Chubu & Kanto & Tohoku & Hokkaido & \quad{Total} \\[-0.1em]
    \midrule
Apr & 51 & 47 & 161 & 100 & 342 & 31 & 24 & \quad756 \tabularnewline
May & 67 & 62 & 127 & 107 & 300 & 47 & 34 & \quad744 \tabularnewline
Jun & 56 & 49 & 140 & 114 & 339 & 38 & 34 & \quad770 \tabularnewline
Jul & 55 & 46 & 132 & 111 & 341 & 31 & 45 & \quad761 \tabularnewline
Aug & 55 & 48 & 145 & 123 & 332 & 37 & 28 & \quad768 \tabularnewline\
Sep & 66 & 64 & 147 & 111 & 302 & 50 & 32 & \quad772 \tabularnewline\
Oct & 54 & 54 & 167 & 108 & 322 & 45 & 40 & \quad790 \tabularnewline\
Nov & 66 & 66 & 149 & 115 & 311 & 34 & 34 & \quad775 \tabularnewline\ \\[-1.4em] 
	\midrule \\[-1.5em] 
Total & 470 & 436 & 1,168 & 889 & 2,589 & 313 & 271 & \quad6,136 \tabularnewline
    \bottomrule
\end{tabularx}
}
\end{threeparttable}
\quad\\[1em]
\begin{flushleft}
{\footnotesize
\qquad \textit{Source:}
Created by the author, based on the \citek{NISTEP-data:2012app} data.
}
\end{flushleft}
%%==================================================
\quad\\
\vfill
}

{\clearpage%
%%==================================================
\vspace*{-3em}
\centering
\begin{threeparttable}
\caption{\textbf{Socio-demographic characteristics of respondents.}}
\label{tab:demog}
{\footnotesize
    \begin{tabular}{l @{\hspace{2.5cm}} rrrr @{\hspace{1.5cm}} rrrr}
    \toprule\\[-2em]
    $n=6,136$ & \multicolumn{4}{l}{Valid raw data} & \multicolumn{4}{l}{After weighting}  \\[-0.2em]
    \midrule
    {\bf Gender} & \multicolumn{2}{l}{0 = \textit{Male}}& \multicolumn{2}{l}{1 = \textit{Female}}& \multicolumn{2}{l}{0 = \textit{Male}}& \multicolumn{2}{l}{1 = \textit{Female}} \\
    \multicolumn{1}{l}{} & 3,055  & (49.8\%) & 3,081   & (50.2\%) & 3,075   & (50.1\%) & 3,061   & (49.9\%) \\[3mm]
    {\bf Age}   &  \multicolumn{4}{l}{${\rm Mean}=40.0$, ~${\rm S.D.}=16.0$}   & \multicolumn{4}{l}{${\rm Mean}=43.4$, ~${\rm S.D.}=14.8$}\\
    \multicolumn{1}{l}{\quad \textit{15--19}} & 501    & (8.2\%) & 510    & (8.3\%) & 215    & (3.5\%) & 209    & (3.4\%) \\
    \multicolumn{1}{l}{\quad \textit{20--29}} & 497    & (8.1\%) & 512    & (8.3\%) & 472    & (7.7\%) & 452    & (7.4\%) \\
    \multicolumn{1}{l}{\quad \textit{30--39}} & 510    & (8.3\%) & 520    & (8.5\%) & 607    & (9.9\%) & 590    & (9.6\%) \\
    \multicolumn{1}{l}{\quad \textit{40--49}} & 514    & (8.4\%) & 514    & (8.4\%) & 619    & (10.1\%) & 608    & (9.9\%) \\
    \multicolumn{1}{l}{\quad \textit{50--59}} & 522    & (8.5\%) & 515    & (8.4\%) & 540    & (8.8\%) & 544    & (8.9\%) \\
    \multicolumn{1}{l}{\quad \textit{60--69}} & 511    & (8.3\%) & 510    & (8.3\%) & 622    & (10.1\%) & 659    & (10.7\%) \\[3mm]
    {\bf Education} &  \multicolumn{4}{l}{${\rm Mean}=0.45$, ~${\rm S.D.}=0.50$}   & \multicolumn{4}{l}{${\rm Mean}=0.46$, ~${\rm S.D.}=0.50$}\\
    \multicolumn{1}{l}{\quad 1 = \textit{High}\tnote{a}} & 1,712 & (27.9\%) & 1,069 & (17.4\%) & 1,779   & (29.0\%) & 1,010 & (16.5\%) \\
    \multicolumn{1}{l}{\quad 0 = \textit{Low}\tnote{b}} & 1,328 & (21.6\%) & 1,997 & (32.5\%) & 1,280   & (20.9\%) & 2,036 & (33.2\%) \\
    \multicolumn{1}{l}{\quad \textit{Other}} & 15     & (0.2\%) & 15     & (0.2\%) & 16     & (0.3\%) & 14     & (0.2\%) \\[2mm]
    {\bf Marital status} &  \multicolumn{4}{l}{${\rm Mean}=0.57$, ~${\rm S.D.}=0.50$}   & \multicolumn{4}{l}{${\rm Mean}=0.65$, ~${\rm S.D.}=0.48$}\\
    \multicolumn{1}{l}{\quad 1 = \textit{Married}} & 1,584 & (25.8\%) & 1,910 & (31.1\%) & 1,825   & (29.7\%) & 2,185 & (35.6\%) \\
    \multicolumn{1}{l}{\quad 0 = \textit{Otherwise}} & 1,471 & (24.0\%) & 1,171 & (19.1\%) & 1,251   & (20.4\%) & 876 & (14.3\%) \\[3mm]
    {\bf Income}\tnote{c} &  \multicolumn{4}{l}{${\rm Mean}=5.1$, ~${\rm S.D.}=3.1$}   & \multicolumn{4}{l}{${\rm Mean}=5.2$, ~${\rm S.D.}=3.1$}\\
    \multicolumn{1}{l}{\quad 1 = \textit{$<$\textyen1,000}} & 335    & (5.5\%) & 332    & (5.4\%) & 287    & (4.7\%) & 266    & (4.3\%) \\
    \multicolumn{1}{l}{\quad 2 = \textit{\textyen1,000--1,999}} & 307    & (5.0\%) & 295    & (4.8\%) & 307    & (5.0\%) & 292    & (4.8\%) \\
    \multicolumn{1}{l}{\quad 3 = \textit{\textyen2,000--2,999}} & 375    & (6.1\%) & 426    & (6.9\%) & 390    & (6.4\%) & 432    & (7.0\%) \\
    \multicolumn{1}{l}{\quad 4 = \textit{\textyen3,000--3,999}} & 433    & (7.1\%) & 451    & (7.4\%) & 439    & (7.1\%) & 453    & (7.4\%) \\
    \multicolumn{1}{l}{\quad 5 = \textit{\textyen4,000--4,999}} & 317    & (5.2\%) & 376    & (6.1\%) & 326    & (5.3\%) & 387    & (6.3\%) \\
    \multicolumn{1}{l}{\quad 6 = \textit{\textyen5,000--5,999}} & 247    & (4.0\%) & 236    & (3.8\%) & 258    & (4.2\%) & 250    & (4.1\%) \\
    \multicolumn{1}{l}{\quad 7 = \textit{\textyen6,000--6,999}} & 172    & (2.8\%) & 195    & (3.2\%) & 177    & (2.9\%) & 200    & (3.3\%) \\
    \multicolumn{1}{l}{\quad 8 = \textit{\textyen7,000--7,999}} & 163    & (2.7\%) & 157    & (2.6\%) & 169    & (2.8\%) & 166    & (2.7\%) \\
    \multicolumn{1}{l}{\quad 9 = \textit{\textyen8,000--8,999}} & 113    & (1.8\%) & 96    & (1.6\%) & 121    & (2.0\%) & 96    & (1.6\%) \\
    \multicolumn{1}{l}{\quad 10 = \textit{\textyen9,000--9,999}} & 140    & (2.3\%) & 131    & (2.1\%) & 142    & (2.3\%) & 128    & (2.1\%) \\
    \multicolumn{1}{l}{\quad 11 = \textit{\textyen10,000--11,999}} & 97     & (1.6\%) & 90    & (1.5\%) & 102     & (1.7\%) & 90    & (1.5\%) \\
    \multicolumn{1}{l}{\quad 12 = \textit{\textyen12,000--14,999}} & 92    & (1.5\%) & 64    & (1.0\%) & 95    & (1.6\%) & 65    & (1.1\%) \\
    \multicolumn{1}{l}{\quad 13 = \textit{\textyen15,000--19,999}} & 36    & (0.6\%) & 28     & (0.5\%) & 37     & (0.6\%) & 29     & (0.5\%) \\
    \multicolumn{1}{l}{\quad 14 = \textit{$>$\textyen20,000}} & 20    & (0.3\%) & 13     & (0.2\%) & 19     & (0.3\%) & 12     & (0.2\%) \\
    \multicolumn{1}{l}{\quad \textit{Don't know}} & 208    & (3.4\%) & 191    & (3.1\%) & 206    & (3.4\%) & 193    & (3.2\%) \\
    \bottomrule
    \end{tabular}
}
\begin{tablenotes}[para,flushleft]
\linespread{1.6}\scriptsize
\vspace{-0.5em}
\item[a] College, university and graduate school.\\[-2mm]
\item[b] Junior high school, high school, specialised training college, vocational school, college of technology, and junior college.\\[-2mm]
\item[c] 000/999 omitted. For example, category \q{3} represents the income bracket \textyen2,000,000--2,999,999.
\end{tablenotes}
\end{threeparttable}
\quad\\[1em]
\begin{flushleft}
{\footnotesize
\qquad \textit{Source:}
Created by the author, based on the \citek{NISTEP-data:2012app} data.
}
\end{flushleft}
%%==================================================
\quad\\
\vfill
}

{\clearpage%
%%==================================================
\vspace*{-3em}
\centering
\begin{threeparttable}
\caption{\textbf{Responses to survey questions about public attitudes towards S\&T policymaking.}}
\label{tab:keyQscicom}
{\footnotesize
    \begin{tabular}{p{13em}c@{\hspace{1cm}}c@{\hspace{1cm}}c@{\hspace{1cm}}c@{\hspace{0.5cm}}c}
    \toprule\\[-2em]
    {}& \multicolumn{1}{l}{1} & \multicolumn{1}{l}{2} & \multicolumn{1}{l}{3} & \multicolumn{1}{l}{4} & \multicolumn{1}{l}{Don't know} \\[-0.2em]
    \midrule
    \multicolumn{6}{l}{\textbf{Interest in S\&T} (\q{\texttt{interest}})}  \\
    \multicolumn{6}{l}{~~\it To what extent are you interested in S\&T news and issues?} \\
    \multicolumn{6}{r}{\footnotesize \quad [$1=\text{\q{not interested at all}}, 2=\text{\q{quite uninterested}}, 3=\text{\q{quite interested}}, 4=\text{\q{very interested}}$]} \\[1mm]
    {} &            297  &         1,375  &        3,208  &         1,256  & \multicolumn{1}{c}{$-$} \\[-0.4em]
    {} & (4.8\%) & (22.4\%) & (52.3\%) & (20.5\%) & \multicolumn{1}{c}{$-$} \\[1em]
    \multicolumn{6}{l}{\textbf{Delegation to the experts} (\q{\texttt{delegate}})}  \\
    \multicolumn{6}{l}{~~\it `Decisions about the direction of R\&D in S\&T should be made by trained and experienced experts}  \\[-1mm]
    \multicolumn{6}{l}{\it \quad who are highly knowledgeable'}  \\
    \multicolumn{6}{r}{\footnotesize \quad [$1=\text{\q{strongly disagree}}, 2=\text{\q{tend to disagree}}, 3=\text{\q{tend to agree}}, 4=\text{\q{strongly agree}}$]}  \\[1mm]
    {} &            489  &         1,752  &        2,449  &            433  &         1,014  \\[-0.4em]
    {} & (8.0\%) & (28.5\%) & (39.9\%) & (7.1\%) & (16.5\%) \\[1em]
    \multicolumn{6}{l}{\textbf{Participation of the general public} (\q{\texttt{participate}})}  \\
    \multicolumn{6}{l}{~~\it `Where R\&D has a significant social impact, the general public should participate in some way in} \\[-1mm]
    \multicolumn{6}{l}{\it \quad making decisions about R\&D'.} \\
    \multicolumn{6}{r}{\footnotesize \quad [$1=\text{\q{strongly disagree}}, 2=\text{\q{tend to disagree}}, 3=\text{\q{tend to agree}}, 4=\text{\q{strongly agree}}$]}  \\[1mm]
    {} &              110  &            664  &        3,346  &          1,126  &             891  \\[-0.4em]
    {} & (1.8\%) & (10.8\%) & (54.5\%) & (18.3\%) & (14.5\%) \\
    \bottomrule
    \end{tabular}
}
%\begin{tablenotes}[para,flushleft]
%\end{tablenotes}
\end{threeparttable}
\quad\\[1em]
\begin{flushleft}
{\footnotesize
\qquad \textit{Source:}
Created by the author, based on the \citek{NISTEP-data:2012app} data.
}
\end{flushleft}
%%==================================================
\quad\\
\vfill
}

{\clearpage%
%%==================================================
\vspace*{-3em}
\centering
\begin{threeparttable}
\caption{\textbf{Responses to survey questions about the public's belief in S\&T-based society.}}
\label{tab:keyQbelief}
{\footnotesize
    \begin{tabular}{p{13em}c@{\hspace{1cm}}c@{\hspace{1cm}}c@{\hspace{1cm}}c@{\hspace{1cm}}c}
    \toprule\\[-2em]
          & \multicolumn{1}{l}{1}     & \multicolumn{1}{l}{2}     & \multicolumn{1}{l}{3}     & \multicolumn{1}{l}{4}     & \multicolumn{1}{l}{5} \\[-0.2em]
    \midrule
    \multicolumn{6}{l}{\textbf{Belief in S\&T-based society} (\q{\texttt{belief}})} \\
    \multicolumn{6}{l}{~~\it `To what extent do you believe that it is important \dots ?'}  \\
    \multicolumn{6}{r}{\footnotesize \quad [$1=\text{\q{not at all}}, 3=\text{\q{to some extent}}, 5=\text{\q{very much}}$]}  \\[2mm]
    \multicolumn{6}{l}{($S_{1}$) \textit{\q{\dots to develop researchers with creativity and high ability levels}}} \\
~(\q{\texttt{researcher}})         &               83  &             253  &          1,889  &          1,727  &          2,184  \\[-0.4em]
          & (1.4\%) & (4.1\%) & (30.8\%) & (28.1\%) & (35.6\%) \\[1em]
    \multicolumn{6}{l}{($S_{2}$) \textit{\q{\dots to actively support universities' research and education programmes}}}  \\
~(\q{\texttt{univsupport}})          &             112  &             358  &          2,140  &          1,828  &          1,698  \\[-0.4em]
          & (1.8\%) & (5.8\%) & (34.9\%) & (29.8\%) & (27.7\%) \\[1em]
    \multicolumn{6}{l}{($S_{3}$) \textit{\q{\dots that the government and public agencies conduct R\&D to solve social challenges}}} \\
~(\q{\texttt{topdown}})       &             150  &             454  &          2,199  &          1,829  &          1,504  \\[-0.4em]
          & (2.4\%) & (7.4\%) & (35.8\%) & (29.8\%) & (24.5\%) \\[1em]
    \multicolumn{6}{l}{($S_{4}$) \textit{\q{\dots to promote the practical application of research results that bring about innovation}}}  \\
~(\q{\texttt{innovation}})       &               82  &             296  &          1,895  &          1,932  &          1,931  \\[-0.4em]
          & (1.3\%) & (4.8\%) & (30.9\%) & (31.5\%) & (31.5\%) \\
    \bottomrule
    \end{tabular}
}
%\begin{tablenotes}[para,flushleft]
%\end{tablenotes}
\end{threeparttable}
\quad\\[1em]
\begin{flushleft}
{\footnotesize
\qquad \textit{Source:}
Created by the author, based on the \citek{NISTEP-data:2012app} data.
}
\end{flushleft}
%%==================================================
\quad\\
\vfill
}

{\clearpage%
%%==================================================
\vspace*{-3em}
\centering
\begin{threeparttable}
\caption{\textbf{Responses to survey questions about the public's socio-scientific knowledge.}}
\label{tab:keyQknowledge}
{\footnotesize
    \begin{tabular}{p{20em}c@{\hspace{1.6cm}}c@{\hspace{1.1cm}}c}
    \toprule\\[-2em]
          & Correct & Incorrect & Don't know \\[-0.2em]
    \midrule
    \multicolumn{4}{l}{\textbf{Socio-scientific knowledge} (\q{\texttt{knowledge}})} \\
    \multicolumn{4}{r}{{\footnotesize \quad [True/False question: $\text{T}=\text{\q{True}}, \text{F}=\text{\q{False}}$]}} \\[2mm]
    \multicolumn{4}{l}{($Q_{1}$) \textit{\q{All radioactive substances are man-made or artificial}.} [F]}  \\
~(\q{\texttt{allmanmade}})          &          4,085  &             789  &          1,263  \\[-0.4em]
          & (66.6\%) & (12.9\%) & (20.6\%) \\[1em]
    \multicolumn{4}{l}{($Q_{2}$) \textit{\q{In 2009, Japan generated approximately 30\% of its domestic power from nuclear energy}.} [T]} \\
~(\q{\texttt{nukpower30}})        &          3,135  &             812  &          2,189  \\[-0.4em]
          & (51.1\%) & (13.2\%) & (35.7\%) \\[1em]
    \multicolumn{4}{l}{($Q_{3}$) \textit{\q{Nuclear power produces no CO${}_{\it 2}$ emissions during electricity generating operations}.} [T]}  \\
~(\q{\texttt{noCO2emit}})        &          3,415  &          1,029  &          1,692  \\[-0.4em]
          & (55.7\%) & (16.8\%) & (27.6\%) \\[1em]
    \multicolumn{4}{l}{($Q_{4}$) \textit{\q{People are exposed to natural radiation in their daily life}.} [T]}  \\
~(\q{\texttt{exposedaily}})       &          5,497  &             117  &              522  \\[-0.4em]
          & (89.6\%) & (1.9\%) & (8.5\%) \\
    \bottomrule
    \end{tabular}
}
%\begin{tablenotes}[para,flushleft]
%\end{tablenotes}
\end{threeparttable}
\quad\\[1em]
\begin{flushleft}
{\footnotesize
\qquad \textit{Source:}
Created by the author, based on the \citek{NISTEP-data:2012app} data.
}
\end{flushleft}
%%==================================================
\quad\\
\vfill
}

{\clearpage%
%%==================================================
\vspace*{-3em}
\centering
\begin{threeparttable}
\caption{\textbf{Responses to survey questions about about the public’s trust in S\&T experts.}}
\label{tab:keyQtrust}
{\footnotesize
    \begin{tabular}{p{13em}c@{\hspace{1cm}}c@{\hspace{1cm}}c@{\hspace{1cm}}c@{\hspace{0.5cm}}c}
    \toprule\\[-2em]
    {}& \multicolumn{1}{l}{1} & \multicolumn{1}{l}{2} & \multicolumn{1}{l}{3} & \multicolumn{1}{l}{4} & \multicolumn{1}{l}{Don't know} \\[-0.2em]
    \midrule
    \multicolumn{6}{l}{\textbf{Trust in scientists} (\q{\texttt{trust\_sci}})}   \\
    \multicolumn{6}{l}{~~\q{\it Do you find scientists' statements trustworthy?}}  \\
     \multicolumn{6}{r}{\footnotesize \quad [$1=\text{\q{very untrustworthy}}, 2=\text{\q{quite untrustworthy}}, 3=\text{\q{quite trustworthy}}, 4=\text{\q{very trustworthy}}$]}  \\[1mm]
    {} &            207  &            835  &        3,489  &            380  &        1,225  \\[-0.4em]
    {} & (3.4\%) & (13.6\%) & (56.9\%) & (6.2\%) & (20.0\%) \\[1em]
    \multicolumn{6}{l}{\textbf{Trust in engineers} (\q{\texttt{trust\_eng}})}   \\
    \multicolumn{6}{l}{~~\q{\it Do you find engineers' statements trustworthy?}}  \\
     \multicolumn{6}{r}{\footnotesize \quad [$1=\text{\q{very untrustworthy}}, 2=\text{\q{quite untrustworthy}}, 3=\text{\q{quite trustworthy}}, 4=\text{\q{very trustworthy}}$]}  \\[1mm]
    {} &             128  &            507  &        3,679  &            893  &            930  \\[-0.4em]
    {} & (2.1\%) & (8.3\%) & (60.0\%) & (14.6\%) & (15.2\%) \\
    \bottomrule
    \end{tabular}
}
%\begin{tablenotes}[para,flushleft]
%\end{tablenotes}
\end{threeparttable}
\quad\\[1em]
\begin{flushleft}
{\footnotesize
\qquad \textit{Source:}
Created by the author, based on the \citek{NISTEP-data:2012app} data.
}
\end{flushleft}
%%==================================================
\quad\\
\vfill
}

{\clearpage%
%%==================================================
\vspace*{-3em}
\centering
\begin{threeparttable}
\caption{\textbf{Summary statistics (top) and results from confirmatory factor analysis (bottom) for \q{belief in S\&T-based society}.}}
\label{tab:FA:belief}
{\footnotesize
    \begin{tabular}{l@{\hspace{1.5cm}}c@{\hspace{1cm}}c@{\hspace{1cm}}c@{\hspace{0.6cm}}c@{\hspace{0.6cm}}c@{\hspace{1.5cm}}c}
    \toprule\\[-2em]
    \multirow{2}[0]{*}{Indicators} & \multirow{2}[0]{*}{Mean} & \multirow{2}[0]{*}{S.D.} & \multicolumn{3}{l}{Polychoric correlation} & {Cronbach's alpha}  \\[-0.4em]
          &       &       & {\tt S1}    & {\tt S2}    & {\tt S3}    & (${\rm overall}=0.89$) \\[-0.2em]
    \midrule
    {\tt S1}: {\tt researcher} & 3.92  & 0.97  & $-$     &       &       & 0.86 \\
    {\tt S2}: {\tt univsupport} & 3.76  & 0.98  & 0.80  & $-$     &       & 0.84 \\
    {\tt S3}: {\tt topdown} & 3.67  & 1.00  & 0.65  & 0.75  & $-$     & 0.86 \\
    {\tt S4}: {\tt innovation} & 3.87  & 0.96  & 0.74  & 0.73  & 0.75  & 0.85 \\
    \bottomrule
    \end{tabular}
\vspace{1.0cm}
    \begin{tabular}{lll@{\hspace{1cm}}r@{\hspace{1cm}}r@{\hspace{1cm}}r@{\hspace{1cm}}r}
    \toprule\\[-2em]
          &       &       & {Loading} & {S.E.} & {$z$-value} & {$p$-value} \\[-0.2em]
    \midrule
   {\tt S1}: {\tt researcher} & $\longleftarrow$    & {\tt belief} & 1.00  & \multicolumn{3}{c}{(constrained)} \\[-1mm]
          &       & {\tt constant} & 3.90  & 0.012 & 312.4 & 0.000 \\
   {\tt S2}: {\tt univsupport} & $\longleftarrow$    & {\tt belief} & 1.07  & 0.014 & 74.0  & 0.000 \\[-1mm]
          &       & {\tt constant} & 3.76  & 0.013 & 298.9 & 0.000 \\
   {\tt S3}: {\tt topdown} & $\longleftarrow$    & {\tt belief} & 1.01  & 0.016 & 64.4  & 0.000 \\[-1mm]
          &       & {\tt constant} & 3.67  & 0.013 & 286.5 & 0.000 \\
   {\tt S4}: {\tt innovation} & $\longleftarrow$    & {\tt belief} & 1.00  & 0.015 & 68.0  & 0.000 \\[-1mm]
          &       & {\tt constant} & 3.86  & 0.012 & 313.9 & 0.000 \\
    \bottomrule
    \end{tabular}
}
\vspace{0.5em}
\begin{tablenotes}[para,flushleft]
\linespread{1.1}\scriptsize
\textit{Note:} 
$n=6,136$; 
Likelihood ratio (LR) $\chi^2(2)=504.8$, $p<0.001$; 
Standardised root mean squared residual $\text{(SRMR)}=0.025$; 
Coefficient of determination $\text{(CD)}=0.891$.
\end{tablenotes}
\end{threeparttable}
\vspace{5mm}
%%==================================================
\quad\\
\vfill
}

{\clearpage%
%%==================================================
\vspace*{-3em}
\centering
\begin{threeparttable}
\caption{\textbf{Summary statistics (top) and results from confirmatory factor analysis (bottom) for \q{scientific knowledge}.}}
\label{tab:FA:knowledge}
{\footnotesize
    \begin{tabular}{l@{\hspace{1.5cm}}c@{\hspace{1cm}}c@{\hspace{1cm}}c@{\hspace{0.6cm}}c@{\hspace{0.6cm}}c@{\hspace{1.5cm}}c}
    \toprule\\[-2em]
    \multirow{2}[0]{*}{Indicators} & \multirow{2}[0]{*}{Mean} & \multirow{2}[0]{*}{S.D.} & \multicolumn{3}{l}{Polychoric correlation} & {KR20}  \\[-0.4em]
          &       &       & {\tt Q1}    & {\tt Q2}    & {\tt Q3}    & (${\rm overall}=0.63$) \\[-0.2em]
    \midrule
    {\tt Q1}: {\tt allmanmade}		& 0.67  & 0.47  & $-$     &       &       & 0.56 \\
    {\tt Q2}: {\tt nukpower30\%}	& 0.51  & 0.50  & 0.38  & $-$     &       & 0.58 \\
    {\tt Q3}: {\tt noCO2emit}		& 0.56  & 0.50  & 0.44  & 0.49  & $-$     & 0.56 \\
    {\tt Q4}: {\tt exposedaily}		& 0.90  & 0.31  & 0.71  & 0.59  & 0.63  & 0.54 \\
    \bottomrule
    \end{tabular}
\vspace{1.0cm}
    \begin{tabular}{lll@{\hspace{1cm}}r@{\hspace{1cm}}r@{\hspace{1cm}}r@{\hspace{1cm}}r}
    \toprule\\[-2em]
          &       &       & {Loading} & {S.E.} & {$z$-value} & {$p$-value} \\[-0.2em]
    \midrule
   {\tt Q1}: {\tt allmanmade} & $\longleftarrow$    & {\tt knowledge} & 1.00  & \multicolumn{3}{c}{(constrained)} \\[-1mm]
          &       & {\tt constant} & 0.653  & 0.0061 & 107.5 & 0.000 \\
   {\tt Q2}: {\tt nukpower30\%} & $\longleftarrow$   & {\tt knowledge} & 0.897  & 0.0396 & 22.6  & 0.000 \\[-1mm]
          &       & {\tt constant} & 0.503  & 0.0064 & 78.8 & 0.000 \\
   {\tt Q3}: {\tt noCO2emit} & $\longleftarrow$    & {\tt knowledge} & 0.974  & 0.0412 & 23.7  & 0.000 \\[-1mm]
          &       & {\tt constant} & 0.545  & 0.0064 & 85.7 & 0.000 \\
   {\tt Q4}: {\tt exposedaily} & $\longleftarrow$    & {\tt knowledge} & 0.726  & 0.0265 & 27.4  & 0.000 \\[-1mm]
          &       & {\tt constant} & 0.884  & 0.0041 & 216.4 & 0.000 \\
    \bottomrule
    \end{tabular}
}
\vspace{0.5em}
\begin{tablenotes}[para,flushleft]
\linespread{1.1}\scriptsize
\textit{Note:} 
$n=6,136$; 
Likelihood ratio (LR) $\chi^2(2)=97.4$, $p<0.001$; 
Standardised root mean squared residual $\text{(SRMR)}=0.024$; 
Coefficient of determination $\text{(CD)}=0.638$.
\end{tablenotes}
\end{threeparttable}
%%==================================================
\quad\\
\vfill
}

{\clearpage%
%%==================================================
\vspace*{-3em}
\centering
\begin{threeparttable}
\caption{\textbf{Results from multinomial regression analyses.}}
\label{tab:mlogit}
{\footnotesize
{\tabcolsep = 3mm
\begin{tabular}{l*{4}{D{.}{.}{-1}}}
\toprule
{\bfseries (a) Attentive}            &\multicolumn{1}{c}{\qquad Model 1}&\multicolumn{1}{c}{\qquad Model 2}&\multicolumn{1}{c}{\qquad Model 3}&\multicolumn{1}{c}{\qquad Model 4}\\
\midrule
{\tt female}	&-0.200^{*}	&-0.114	&-0.0539	&-0.0817	\\[-2mm]
	&(0.0944)	&(0.104)	&(0.105)	&(0.105)	\\
\addlinespace
{\tt age}	&0.0169^{***}	&0.0205^{***}	& 0.0173^{***}	& 0.0173^{***}	\\[-2mm]
	&(0.00360)		&(0.00402)		&(0.00407)		&(0.00407)	\\
\addlinespace
{\tt educ}  &       0.351^{***}&       0.379^{***}&       0.334^{**}&       0.334^{**}	\\[-2mm]
            &      (0.0919)         &      (0.102)         &               (0.103)         &      (0.103)	\\
\addlinespace
{\tt married}   &     -0.0754         &     -0.103         &     -0.0740         &     -0.0722	\\[-2mm]
            &     (0.114)         &     (0.128)         &     (0.129)         &     (0.129)	\\
\addlinespace
{\tt income}      &     -0.0180         &    -0.00427         &    -0.00825         &    -0.00912	\\[-2mm]
            &     (0.0161)         &     (0.0176)         &     (0.0176)         &     (0.0176)	\\
\addlinespace
{\tt interest}  &       1.536^{***}&       1.762^{***}&       1.728^{***}&       1.681^{***}	\\[-2mm]
            &     (0.0876)                       &     (0.0987)         &     (0.100)         &     (0.101)	\\
\addlinespace
{\tt trust\_sci}    &                     &      -0.621^{***}&      -0.624^{***}&      -0.665^{***}	\\[-2mm]
            &                               &     (0.100)                  &     (0.101)         &     (0.102)	\\
\addlinespace
{\tt trust\_eng}    &                     &     0.0264         &     0.00375         &     -0.0156	\\[-2mm]
            &                     &      (0.104)         &     (0.104)         &     (0.105)	\\
\addlinespace
{\tt knowledge} &                     &                     &       1.342^{***}&       1.195^{***}	\\[-2mm]
            &                     &                     &      (0.301)         &      (0.302)	\\
\addlinespace
{\tt belief} &                     &                     &                     &       0.231^{**}	\\[-2mm]
            &                     &                     &                     &      (0.0761)	\\
%\addlinespace
%{\tt constant}      &      -6.781^{***}&      -6.217^{***}&      -6.047^{***}&      -5.763^{***}	\\[-2mm]
%            &    (0.372)         &    (0.464)         &    (0.471)         &    (0.480)	\\
\multicolumn{5}{c}{\footnotesize \emph{(continued to next page)}}\\
\midrule
\end{tabular}
}
}
%\begin{tablenotes}[para,flushleft]
%\end{tablenotes}
\end{threeparttable}
%%==================================================
\quad\\
\vfill
}

{\clearpage%
%%==================================================
\vspace*{-3em}
\centering
\begin{threeparttable}\ContinuedFloat
\caption{\textbf{Results from multinomial regression analyses.} \emph{(Cont.)}}
\label{tab:mlogit}
{
\footnotesize
{\tabcolsep = 3mm
\begin{tabular}{l*{4}{D{.}{.}{-1}}}
\toprule
{\bfseries (b) Residual}            &\multicolumn{1}{c}{\qquad Model 1}&\multicolumn{1}{c}{\qquad Model 2}&\multicolumn{1}{c}{\qquad Model 3}&\multicolumn{1}{c}{\qquad Model 4}	\\
\midrule
{\tt female}    &      -0.125&      -0.150^{*}&      -0.156^{*}&      -0.126	\\[-2mm]
            &     (0.0652)         &     (0.0758)         &     (0.0761)         &     (0.0766)	\\
\addlinespace
{\tt age}         &      0.0124^{***}&      0.0132^{***}&      0.0137^{***}&      0.0142^{***}	\\[-2mm]
            &      (0.00259)         &      (0.00298)         &      (0.00302)         &      (0.00304)	\\
\addlinespace
{\tt educ}  &       0.0343 &       0.0849           &       0.0939 &       0.106	\\[-2mm]
            &      (0.0639)         &      (0.0733)         &      (0.0739)         &      (0.0741)	\\
\addlinespace
{\tt married}   &       0.0418         &       0.153         &       0.151         &       0.141	\\[-2mm]
            &      (0.0808)         &      (0.0950)         &      (0.0949)         &      (0.0954)	\\
\addlinespace
{\tt income}      &     0.00460         &      0.0129         &      0.0137         &      0.0142	\\[-2mm]
            &      (0.0113)         &      (0.0129)         &      (0.0129)         &      (0.0129)	\\
\addlinespace
{\tt interest}  &      -0.419^{***}&      -0.286^{***}&      -0.277^{***}&      -0.239^{***}	\\[-2mm]
            &     (0.0429)         &     (0.0523)         &     (0.0536)         &     (0.0545)	\\
\addlinespace
{\tt trust\_sci}    &                     &      -0.411^{***}&      -0.410^{***}&      -0.372^{***}	\\[-2mm]
            &                     &     (0.0734)         &     (0.0733)         &     (0.0743)	\\
\addlinespace
{\tt trust\_eng}    &                     &      -0.265^{***}&      -0.260^{***}&      -0.233^{**}	\\[-2mm]
            &                     &     (0.0733)         &     (0.0735)         &     (0.0741)	\\
\addlinespace
{\tt knowledge} &                     &                     &       -0.167         &       -0.0755	\\[-2mm]
            &                     &                     &      (0.190)         &      (0.192)	\\
\addlinespace
{\tt belief} &                     &                     &                     &      -0.197^{***}	\\[-2mm]
            &                     &                     &                     &     (0.0516)	\\
\addlinespace
%\addlinespace
%{\tt constant}      &      0.643^{***}  &       1.480^{***}&       1.421^{***}&       1.111^{***}	\\[-2mm]
%            &     (0.189)         &      (0.277)         &      (0.284)         &      (0.295)	\\
\midrule
$N$	&\multicolumn{1}{c}{5,711}	&\multicolumn{1}{c}{4,505}	&\multicolumn{1}{c}{4,505}	&\multicolumn{1}{c}{4,505}	\\
Log-likelihood &&&&{}\\
\quad Model	&-5490.86	&-4296.56	&-4283.75	&-4263.78	\\
\quad Intercept-only	&-6090.21	&-4879.93	&-4879.93	&-4879.93	\\
\quad Likelihood ratio	&1198.68	&1166.74	&1192.37	&1232.30	\\
\quad $p$-value	&\multicolumn{1}{c}{$<0.001$}	&\multicolumn{1}{c}{$<0.001$}	&\multicolumn{1}{c}{$<0.001$}	&\multicolumn{1}{c}{$<0.001$}	\\
Pseudo-$R^2$ &&&&{}\\
\quad McFadden's	&0.098	&0.120	&0.122	&0.126	\\
\quad Cragg\,\&\,Uhler's	&0.189	&0.228	&0.233	&0.239	\\
\bottomrule
\end{tabular}
}
}
\vspace{0.5em}
\begin{tablenotes}[para,flushleft]
\linespread{1.1}\scriptsize
\textit{Note:} 
Four models (Model 1–-Model 4), corresponding to different specifications of independent variables described in Eq.~(\ref{eq:model1})--(\ref{eq:model4}), are compared. 
For all models, the base category is \q{Interested}. 
The outcome category is \q{Attentive} in part (a) and \q{Residual} in part (b), respectively.
Regression coefficients with robust standard errors in parentheses.
${}^{*}$ $p<0.05$, ${}^{**}$ $p<0.01$, ${}^{***}$ $p<0.001$.
Only the coefficients of particular interest are shown to save space; the variables \q{\texttt{Region}} and \q{\texttt{YearMonth}} are omitted from display.
\end{tablenotes}
\end{threeparttable}
%%==================================================
\quad\\
\vfill
}

\end{document}